\documentclass[twocolumn]{aastex62}
\pdfoutput=1 
\usepackage{amsmath,amstext,amssymb}
\usepackage[T1]{fontenc}
\usepackage{apjfonts} 
\usepackage{subfigure}
\usepackage{xcolor}
\usepackage{soul}
\usepackage{comment}
\usepackage{graphicx}
\setstcolor{red}

\graphicspath{{figures/}}

\newcommand{\de}{{\rm d}}

\shorttitle{Measurement of the Hubble constant from GW170814}
\shortauthors{Soares-Santos, Palmese et al.}

\begin{document}

\title{
First measurement of the Hubble constant from a dark standard siren using the Dark Energy Survey galaxies and the LIGO/Virgo binary--black--hole merger GW170814
}

\author{M.~Soares-Santos}
\affiliation{Department of Physics, Brandeis University, Waltham, MA 02453, USA}

\author{A.~Palmese}
\affiliation{Fermi National Accelerator Laboratory, P. O. Box 500, Batavia, IL 60510, USA}
\correspondingauthor{Antonella Palmese} \email{palmese@fnal.gov}

\author{W.~Hartley}
\affiliation{Department of Physics \& Astronomy, University College London, Gower Street, London, WC1E 6BT, UK}

\author{J.~Annis}
\affiliation{Fermi National Accelerator Laboratory, P. O. Box 500, Batavia, IL 60510, USA}

\author{J.~Garcia-Bellido}
\affiliation{Instituto de Fisica Teorica UAM/CSIC, Universidad Autonoma de Madrid, 28049 Madrid, Spain}

\author{O.~Lahav}
\affiliation{Department of Physics \& Astronomy, University College London, Gower Street, London, WC1E 6BT, UK}

\author{Z.~Doctor}
\affiliation{Department of Physics, University of Chicago, Chicago, IL 60637, USA}
\affiliation{Kavli Institute for Cosmological Physics, University of Chicago, Chicago, IL 60637, USA}

\author{M.~Fishbach}
\affiliation{Kavli Institute for Cosmological Physics, University of Chicago, Chicago, IL 60637, USA}

\author{D.~E.~Holz}
\affiliation{University of Chicago, Chicago, IL 60637, USA}

\author{H.~Lin}
\affiliation{Fermi National Accelerator Laboratory, P. O. Box 500, Batavia, IL 60510, USA}

\author{M.~E.~S.~Pereira}
\affiliation{Department of Physics, Brandeis University, Waltham, MA 02453, USA}

\author{A.~Garcia}
\affiliation{Department of Physics, Brandeis University, Waltham, MA 02453, USA}

\author{K.~Herner}
\affiliation{Fermi National Accelerator Laboratory, P. O. Box 500, Batavia, IL 60510, USA}

\author{R.~Kessler}
\affiliation{Department of Astronomy and Astrophysics, University of Chicago, Chicago, IL 60637, USA}
\affiliation{Kavli Institute for Cosmological Physics, University of Chicago, Chicago, IL 60637, USA}

\author{H.~V.~Peiris}
\affiliation{Department of Physics \& Astronomy, University College London, Gower Street, London, WC1E 6BT, UK}

\author{M.~Sako}
\affiliation{Department of Physics and Astronomy, University of Pennsylvania, Philadelphia, PA 19104, USA}

\author{S.~Allam}
\affiliation{Fermi National Accelerator Laboratory, P. O. Box 500, Batavia, IL 60510, USA}

\author{D.~Brout}
\affiliation{Department of Physics and Astronomy, University of Pennsylvania, Philadelphia, PA 19104, USA}

\author{A.~Carnero Rosell}
\affiliation{Centro de Investigaciones Energ\'eticas, Medioambientales y Tecnol\'ogicas (CIEMAT), Madrid, Spain}
\affiliation{Laborat\'orio Interinstitucional de e-Astronomia - LIneA, Rua Gal. Jos\'e Cristino 77, Rio de Janeiro, RJ - 20921-400, Brazil}

\author{H.~Y.~Chen}
\affiliation{University of Chicago, Chicago, IL 60637, USA}

\author{C.~Conselice}
\affiliation{University of Nottingham, School of Physics and Astronomy, Nottingham NG7 2RD, UK}

\author{J.~deRose}
\affiliation{Kavli Institute for Particle Astrophysics and Cosmology and Department of Physics, Stanford University, Stanford, CA, USA, 94305}
\affiliation{Department of Particle Physics \& Astrophysics, SLAC National Accelerator Laboratory, Menlo Park, CA 94025}

\author{J.~deVicente}
\affiliation{Centro de Investigaciones Energ\'eticas, Medioambientales y Tecnol\'ogicas (CIEMAT), Madrid, Spain}
\author{H.~T.~Diehl}
\affiliation{Fermi National Accelerator Laboratory, P. O. Box 500, Batavia, IL 60510, USA}

\author{M.~S.~S.~Gill}
\affiliation{SLAC National Accelerator Laboratory, Menlo Park, CA 94025, USA}

\author{J.~Gschwend}
\affiliation{Laborat\'orio Interinstitucional de e-Astronomia - LIneA, Rua Gal. Jos\'e Cristino 77, Rio de Janeiro, RJ - 20921-400, Brazil}
\affiliation{Observat\'orio Nacional, Rua Gal. Jos\'e Cristino 77, Rio de Janeiro, RJ - 20921-400, Brazil}


\author{I. Sevilla-Noarbe}
\affiliation{Centro de Investigaciones Energ\'eticas, Medioambientales y Tecnol\'ogicas (CIEMAT), Madrid, Spain}
\author{D.~L.~Tucker}
\affiliation{Fermi National Accelerator Laboratory, P. O. Box 500, Batavia, IL 60510, USA}

\author{R.~Wechsler}
\affiliation{Department of Physics, Stanford University, 382 Via Pueblo Mall, Stanford, CA 94305, USA}
\affiliation{Kavli Institute for Particle Astrophysics \& Cosmology, P. O. Box 2450, Stanford University, Stanford, CA 94305, USA}
\affiliation{SLAC National Accelerator Laboratory, Menlo Park, CA 94025, USA}


\author{E.~Berger}
\affiliation{Harvard-Smithsonian Center for Astrophysics, 60 Garden Street, Cambridge MA 02138, USA}

\author{P.~S.~Cowperthwaite}
\affiliation{The Observatories of the Carnegie Institution for Science, 813 Santa Barbara St., Pasadena,
CA 91101, USA}
\affiliation{Hubble Fellow}

\author{B.~D.~Metzger}
\affiliation{Department of Physics, Columbia University, New York, NY 10025, USA}

\author{P.~K.~G.~Williams}
\affiliation{Harvard-Smithsonian Center for Astrophysics, 60 Garden Street, Cambridge MA 02138, USA}
\affiliation{American Astronomical Society, 1667 K Street NW, Suite 800 Washington, DC 20006, USA}


\author{T.~M.~C.~Abbott}
\affiliation{Cerro Tololo Inter-American Observatory, National Optical Astronomy Observatory, Casilla 603, La Serena, Chile}
\author{F.~B.~Abdalla}
\affiliation{Department of Physics \& Astronomy, University College London, Gower Street, London, WC1E 6BT, UK}
\author{S.~Avila}
\affiliation{Institute of Cosmology and Gravitation, University of Portsmouth, Portsmouth, PO1 3FX, UK}
\author{K.~Bechtol}
\affiliation{LSST, 933 North Cherry Avenue, Tucson, AZ 85721, USA}
\affiliation{Physics Department, 2320 Chamberlin Hall, University of Wisconsin-Madison, 1150 University Avenue Madison, WI  53706-1390}
\author{E.~Bertin}
\affiliation{CNRS, UMR 7095, Institut d'Astrophysique de Paris, F-75014, Paris, France}
\affiliation{Sorbonne Universit\'es, UPMC Univ Paris 06, UMR 7095, Institut d'Astrophysique de Paris, F-75014, Paris, France}
\author{D.~Brooks}
\affiliation{Department of Physics \& Astronomy, University College London, Gower Street, London, WC1E 6BT, UK}
\author{E.~Buckley-Geer}
\affiliation{Fermi National Accelerator Laboratory, P. O. Box 500, Batavia, IL 60510, USA}
\author{D.~L.~Burke}
\affiliation{Kavli Institute for Particle Astrophysics \& Cosmology, P. O. Box 2450, Stanford University, Stanford, CA 94305, USA}
\affiliation{SLAC National Accelerator Laboratory, Menlo Park, CA 94025, USA}

\author{M.~Carrasco~Kind}
\affiliation{Department of Astronomy, University of Illinois at Urbana-Champaign, 1002 W. Green Street, Urbana, IL 61801, USA}
\affiliation{National Center for Supercomputing Applications, 1205 West Clark St., Urbana, IL 61801, USA}
\author{J.~Carretero}
\affiliation{Institut de F\'{\i}sica d'Altes Energies (IFAE), The Barcelona Institute of Science and Technology, Campus UAB, 08193 Bellaterra (Barcelona) Spain}
\author{F.~J.~Castander}
\affiliation{Institut d'Estudis Espacials de Catalunya (IEEC), 08034 Barcelona, Spain}
\affiliation{Institute of Space Sciences (ICE, CSIC),  Campus UAB, Carrer de Can Magrans, s/n,  08193 Barcelona, Spain}

\author{M.~Crocce}
\affiliation{Institut d'Estudis Espacials de Catalunya (IEEC), 08034 Barcelona, Spain"}
\affiliation{Institute of Space Sciences (ICE, CSIC),  Campus UAB, Carrer de Can Magrans, s/n,  08193 Barcelona, Spain}

\author{C.~E.~Cunha}
\affiliation{Kavli Institute for Particle Astrophysics \& Cosmology, P. O. Box 2450, Stanford University, Stanford, CA 94305, USA}
\author{C.~B.~D'Andrea}
\affiliation{Department of Physics and Astronomy, University of Pennsylvania, Philadelphia, PA 19104, USA}
\author{L.~N.~da Costa}
\affiliation{Laborat\'orio Interinstitucional de e-Astronomia - LIneA, Rua Gal. Jos\'e Cristino 77, Rio de Janeiro, RJ - 20921-400, Brazil}
\affiliation{Observat\'orio Nacional, Rua Gal. Jos\'e Cristino 77, Rio de Janeiro, RJ - 20921-400, Brazil}
\author{C.~Davis}
\affiliation{Kavli Institute for Particle Astrophysics \& Cosmology, P. O. Box 2450, Stanford University, Stanford, CA 94305, USA}
\author{S.~Desai}
\affiliation{Department of Physics, IIT Hyderabad, Kandi, Telangana 502285, India}
\author{P.~Doel}
\affiliation{Department of Physics \& Astronomy, University College London, Gower Street, London, WC1E 6BT, UK}
\author{A.~Drlica-Wagner}
\affiliation{Fermi National Accelerator Laboratory, P. O. Box 500, Batavia, IL 60510, USA}
\affiliation{University of Chicago, Chicago, IL 60637, USA}
\author{T.~F.~Eifler}
\affiliation{Department of Astronomy/Steward Observatory, 933 North Cherry Avenue, Tucson, AZ 85721-0065, USA}
\affiliation{Jet Propulsion Laboratory, California Institute of Technology, 4800 Oak Grove Dr., Pasadena, CA 91109, USA}
\author{A.~E.~Evrard}
\affiliation{Department of Astronomy, University of Michigan, Ann Arbor, MI 48109, USA}
\affiliation{Department of Physics, University of Michigan, Ann Arbor, MI 48109, USA}
\author{B.~Flaugher}
\affiliation{Fermi National Accelerator Laboratory, P. O. Box 500, Batavia, IL 60510, USA}
\author{P.~Fosalba}
\affiliation{Institut d'Estudis Espacials de Catalunya (IEEC), 08034 Barcelona, Spain}
\affiliation{Institute of Space Sciences (ICE, CSIC),  Campus UAB, Carrer de Can Magrans, s/n,  08193 Barcelona, Spain}
\author{J.~Frieman}
\affiliation{Fermi National Accelerator Laboratory, P. O. Box 500, Batavia, IL 60510, USA}
\affiliation{Kavli Institute for Cosmological Physics, University of Chicago, Chicago, IL 60637, USA}
\author{E.~Gaztanaga}
\affiliation{Institut d'Estudis Espacials de Catalunya (IEEC), 08034 Barcelona, Spain}
\affiliation{Institute of Space Sciences (ICE, CSIC),  Campus UAB, Carrer de Can Magrans, s/n,  08193 Barcelona, Spain}
\author{D.~W.~Gerdes}
\affiliation{Department of Astronomy, University of Michigan, Ann Arbor, MI 48109, USA}
\affiliation{Department of Physics, University of Michigan, Ann Arbor, MI 48109, USA}
\author{D.~Gruen}
\affiliation{Kavli Institute for Particle Astrophysics \& Cosmology, P. O. Box 2450, Stanford University, Stanford, CA 94305, USA}
\affiliation{SLAC National Accelerator Laboratory, Menlo Park, CA 94025, USA}
\author{R.~A.~Gruendl}
\affiliation{Department of Astronomy, University of Illinois at Urbana-Champaign, 1002 W. Green Street, Urbana, IL 61801, USA}
\affiliation{National Center for Supercomputing Applications, 1205 West Clark St., Urbana, IL 61801, USA}
\author{G.~Gutierrez}
\affiliation{Fermi National Accelerator Laboratory, P. O. Box 500, Batavia, IL 60510, USA}
\author{D.~L.~Hollowood}
\affiliation{Santa Cruz Institute for Particle Physics, Santa Cruz, CA 95064, USA}
\author{B.~Hoyle}
\affiliation{Max Planck Institute for Extraterrestrial Physics, Giessenbachstrasse, 85748 Garching, Germany}
\affiliation{Universit\"ats-Sternwarte, Fakult\"at f\"ur Physik, Ludwig-Maximilians Universit\"at M\"unchen, Scheinerstr. 1, 81679 M\"unchen, Germany}
\author{D.~J.~James}
\affiliation{Harvard-Smithsonian Center for Astrophysics, Cambridge, MA 02138, USA}
\author{T.~Jeltema}
\affiliation{Santa Cruz Institute for Particle Physics, Santa Cruz, CA 95064, USA}
\author{K.~Kuehn}
\affiliation{Australian Astronomical Optics, Macquarie University, North Ryde, NSW 2113, Australia}
\author{N.~Kuropatkin}
\affiliation{Fermi National Accelerator Laboratory, P. O. Box 500, Batavia, IL 60510, USA}
\author{T.~S.~Li}
\affiliation{Fermi National Accelerator Laboratory, P. O. Box 500, Batavia, IL 60510, USA}
\affiliation{Kavli Institute for Cosmological Physics, University of Chicago, Chicago, IL 60637, USA}
\author{M.~Lima}
\affiliation{Departamento de F\'isica Matem\'atica, Instituto de F\'isica, Universidade de S\~ao Paulo, CP 66318, S\~ao Paulo, SP, 05314-970, Brazil}
\affiliation{Laborat\'orio Interinstitucional de e-Astronomia - LIneA, Rua Gal. Jos\'e Cristino 77, Rio de Janeiro, RJ - 20921-400, Brazil}
\author{M.~A.~G.~Maia}
\affiliation{Laborat\'orio Interinstitucional de e-Astronomia - LIneA, Rua Gal. Jos\'e Cristino 77, Rio de Janeiro, RJ - 20921-400, Brazil}
\affiliation{Observat\'orio Nacional, Rua Gal. Jos\'e Cristino 77, Rio de Janeiro, RJ - 20921-400, Brazil}
\author{J.~L.~Marshall}
\affiliation{George P. and Cynthia Woods Mitchell Institute for Fundamental Physics and Astronomy, and Department of Physics and Astronomy, Texas A\&M University, College Station, TX 77843,  USA}
\author{F.~Menanteau}
\affiliation{Department of Astronomy, University of Illinois at Urbana-Champaign, 1002 W. Green Street, Urbana, IL 61801, USA}
\affiliation{National Center for Supercomputing Applications, 1205 West Clark St., Urbana, IL 61801, USA}
\author{R.~Miquel}
\affiliation{Instituci\'o Catalana de Recerca i Estudis Avan\c{c}ats, E-08010 Barcelona, Spain}
\affiliation{Institut de F\'{\i}sica d'Altes Energies (IFAE), The Barcelona Institute of Science and Technology, Campus UAB, 08193 Bellaterra (Barcelona) Spain}
\author{E.~Neilsen}
\affiliation{Fermi National Accelerator Laboratory, P. O. Box 500, Batavia, IL 60510, USA}
\author{R.~L.~C.~Ogando}
\affiliation{Laborat\'orio Interinstitucional de e-Astronomia - LIneA, Rua Gal. Jos\'e Cristino 77, Rio de Janeiro, RJ - 20921-400, Brazil}
\affiliation{Observat\'orio Nacional, Rua Gal. Jos\'e Cristino 77, Rio de Janeiro, RJ - 20921-400, Brazil}
\author{A.~A.~Plazas}
\affiliation{Jet Propulsion Laboratory, California Institute of Technology, 4800 Oak Grove Dr., Pasadena, CA 91109, USA}
\affiliation{Department of Astrophysical Sciences, Princeton University, 4 Ivy Lane, Princeton, NJ 08544}
\author{A.~K.~Romer}
\affiliation{Department of Physics and Astronomy, Pevensey Building, University of Sussex, Brighton, BN1 9QH, UK}
\author{A.~Roodman}
\affiliation{Kavli Institute for Particle Astrophysics \& Cosmology, P. O. Box 2450, Stanford University, Stanford, CA 94305, USA}
\affiliation{SLAC National Accelerator Laboratory, Menlo Park, CA 94025, USA}
\author{E.~Sanchez}
\affiliation{Centro de Investigaciones Energ\'eticas, Medioambientales y Tecnol\'ogicas (CIEMAT), Madrid, Spain}
\author{V.~Scarpine}
\affiliation{Fermi National Accelerator Laboratory, P. O. Box 500, Batavia, IL 60510, USA}
\author{R.~Schindler}
\affiliation{SLAC National Accelerator Laboratory, Menlo Park, CA 94025, USA}
\author{M.~Schubnell}
\affiliation{Department of Physics, University of Michigan, Ann Arbor, MI 48109, USA}
\author{S.~Serrano}
\affiliation{Institut d'Estudis Espacials de Catalunya (IEEC), 08034 Barcelona, Spain}
\affiliation{Institute of Space Sciences (ICE, CSIC),  Campus UAB, Carrer de Can Magrans, s/n,  08193 Barcelona, Spain}
\author{M.~Smith}
\affiliation{School of Physics and Astronomy, University of Southampton,  Southampton, SO17 1BJ, UK}
\author{R.~C.~Smith}
\affiliation{Cerro Tololo Inter-American Observatory, National Optical Astronomy Observatory, Casilla 603, La Serena, Chile}
\author{F.~Sobreira}
\affiliation{Instituto de F\'isica Gleb Wataghin, Universidade Estadual de Campinas, 13083-859, Campinas, SP, Brazil}
\affiliation{Laborat\'orio Interinstitucional de e-Astronomia - LIneA, Rua Gal. Jos\'e Cristino 77, Rio de Janeiro, RJ - 20921-400, Brazil}
\author{E.~Suchyta}
\affiliation{Computer Science and Mathematics Division, Oak Ridge National Laboratory, Oak Ridge, TN 37831}
\author{M.~E.~C.~Swanson}
\affiliation{National Center for Supercomputing Applications, 1205 West Clark St., Urbana, IL 61801, USA}
\author{G.~Tarle}
\affiliation{Department of Physics, University of Michigan, Ann Arbor, MI 48109, USA}
\author{R.~C.~Thomas}
\affiliation{Lawrence Berkeley National Laboratory, 1 Cyclotron Road, Berkeley, CA 94720, USA}
\author{A.~R.~Walker}
\affiliation{Cerro Tololo Inter-American Observatory, National Optical Astronomy Observatory, Casilla 603, La Serena, Chile}
\author{W.~Wester}
\affiliation{Fermi National Accelerator Laboratory, P. O. Box 500, Batavia, IL 60510, USA}
\author{J.~Zuntz}
\affiliation{Institute for Astronomy, University of Edinburgh, Edinburgh EH9 3HJ, UK}

\collaboration{(The DES Collaboration)}

\author{B.~P.~Abbott}
\affiliation{LIGO, California Institute of Technology, Pasadena, CA 91125, USA}
\author{R.~Abbott}
\affiliation{LIGO, California Institute of Technology, Pasadena, CA 91125, USA}
\author{T.~D.~Abbott}
\affiliation{Louisiana State University, Baton Rouge, LA 70803, USA}
\author{S.~Abraham}
\affiliation{Inter-University Centre for Astronomy and Astrophysics, Pune 411007, India}
\author{F.~Acernese}
\affiliation{Universit\`a di Salerno, Fisciano, I-84084 Salerno, Italy}
\affiliation{INFN, Sezione di Napoli, Complesso Universitario di Monte S.Angelo, I-80126 Napoli, Italy}
\author{K.~Ackley}
\affiliation{OzGrav, School of Physics \& Astronomy, Monash University, Clayton 3800, Victoria, Australia}
\author{C.~Adams}
\affiliation{LIGO Livingston Observatory, Livingston, LA 70754, USA}
\author{R.~X.~Adhikari}
\affiliation{LIGO, California Institute of Technology, Pasadena, CA 91125, USA}
\author{V.~B.~Adya}
\affiliation{Max Planck Institute for Gravitational Physics (Albert Einstein Institute), D-30167 Hannover, Germany}
\affiliation{Leibniz Universit\"at Hannover, D-30167 Hannover, Germany}
\author{C.~Affeldt}
\affiliation{Max Planck Institute for Gravitational Physics (Albert Einstein Institute), D-30167 Hannover, Germany}
\affiliation{Leibniz Universit\"at Hannover, D-30167 Hannover, Germany}
% ...

\collaboration{(The LIGO Scientific Collaboration and the Virgo Collaboration)}

\begin{abstract}
    We present a multi--messenger measurement of the Hubble constant $H_0$ using the binary--black--hole merger GW170814 as a standard siren, combined with a photometric redshift catalog from the Dark Energy Survey (DES). The luminosity distance is obtained from the gravitational wave signal detected by the LIGO/Virgo Collaboration (LVC) on 2017 August 14, and the redshift information is provided by the DES Year 3 data. Black--hole mergers such as GW170814 are expected to lack bright electromagnetic emission to uniquely identify their host galaxies and build an object--by--object Hubble diagram. However, they are suitable for a statistical measurement, provided that a galaxy catalog of adequate depth and redshift completion is available. 
    Here we present the first Hubble parameter measurement using a black--hole merger. Our analysis results in $H_0 = 75^{+40}_{-32}~{\rm km~s^{-1}~Mpc^{-1}}$, which is consistent with both SN Ia and CMB measurements of the Hubble constant. The quoted 68\% credible region comprises 60\% of the uniform prior range [20,140] ${\rm km~s^{-1}~Mpc^{-1}}${, and it depends on the assumed prior range. If we take a broader prior of [10,220] ${\rm km~s^{-1}~Mpc^{-1}}$, we find $H_0 = 78^{+ 96}_{-24}~{\rm km~s^{-1}~Mpc^{-1}}$ ($57\%$ of the prior range). Although a weak constraint on the Hubble constant from a single event is expected using the dark siren method, a multifold increase in the LVC event rate is anticipated in the coming years and combinations of many sirens will lead to improved constraints on $H_0$.}

\end{abstract}

\keywords{catalogs --- cosmology: observations --- gravitational waves --- surveys}

\reportnum{DES-2018-0389}
\reportnum{FERMILAB-PUB-18-629-AE}

\section{Introduction}

Unlike most extragalactic distance observables, mergers of neutron star and black hole binary systems are \emph{absolute} distance indicators. Often referred to as ``standard sirens'', they emit gravitational waves (GW) from which the luminosity distance can be inferred without relying on any calibration with respect to another source: the rate of change in frequency gives the system's size and thus the intrinsic amplitude, which is compared against the observed signal amplitude to obtain the distance to the source.  If redshifts are associated with those sirens (in the simplest case, the host galaxy is identified and its redshift is obtained via spectroscopic follow up), a measurement of the present rate of expansion of the Universe $H_0$ can be achieved via the distance--redshift relation. {The use of gravitational wave sources as cosmological probes was first proposed by \citet{schutz}, and recently revisited in several works (e.g. \citealt{2005ApJ...629...15H}).}

For dark energy research, the possibility of measuring $H_0$ directly and independently from other methods is of great interest. Local measurements obtained from type Ia Supernovae (SN Ia) and other distance indicators, as well as the predicted value inferred from the cosmic microwave background at $z\sim 1100$, have achieved remarkable precision of $1-2.5\%$ (e.g. \citealt{riess18,planck18}). They disagree, however, by more than $3\sigma$ and interpreting this tension as evidence for beyond-$\Lambda$CDM dark energy or new physics at the early universe requires new measurements of great precision and accuracy (\citealt{2017NatAs...1E.169F,2018JCAP...09..025M}). Those measurements are one of the greatest challenges faced by current experiments in cosmology because the observables are subject to correlated systematic effects arising from their complex astrophysics. As estimates become more precise, this challenge becomes more severe and the need for novel independent methods becomes more pressing. Those methods, however, are few and hard to come by. One possibility is standard sirens, which remained elusive for almost 30 years, until the {detection of the first gravitational wave event (GW150914; \citealt{GW150914}). The first standard siren-based $H_0$ measurement (\citealt{2017Natur.551...85A})  came with the} discovery of the binary--neutron--star (BNS) merger GW170817 (\citealt{ligobns}) and its associated electromagnetic counterpart (\citealt{MMApaper,marcelle17,arcavi,Coulter1556,lipunov,tanvir,valenti}). Several studies have developed methodologies to infer cosmological parameters from standard sirens and establish their constraining power \citep{schutz,2005ApJ...629...15H,macleod,2010ApJ...725..496N,delpozzo,2013arXiv1307.2638N,2017PhRvD..96j1303N,chen17,2018arXiv180203404F,2018PhRvL.121b1303V,2018arXiv181111723M}. \citet{chen17} predict that we will be able to constrain $H_0$ with 2\% precision within 5 years with standard sirens detected by LIGO/Virgo, while \citet{nair} predict a $\sim 7\%$ measurement with just 25 binary black hole (BBH) events from the Einstein telescope. 

Anticipating that the LIGO/Virgo Collaboration (LVC) network of gravitational wave detectors would eventually achieve sensitivity sufficient to enable standard siren--based measurements, the Dark Energy Survey (DES) collaboration {and external collaborators} launched in 2015 the DES gravitational waves (DESGW) program. DESGW uses DECam to search for optical emission associated with LVC detected mergers and pursues cosmological measurements with standard sirens. In particular, the multi-messenger shared discovery of the neutron--star merger GW170817 and of its {optical kilonova}, resulted in a measurement of $H_0$ (\citealt{2017Natur.551...85A}) that inaugurated the era of siren-based cosmology. We have also performed the most comprehensive searches for {optical emission to} black hole events, including GW150914 (\citealt{2016ApJ...823L..33S}), GW151226 (\citealt{2016ApJ...826L..29C}), and GW170814 (\citealt{doctor}). These events are expected to be dark, although the possibility of optical emission has yet to be observationally excluded. 

Dark sirens can also be used for cosmology using a statistical method, as first proposed in \citet{schutz}. Provided a catalog of potential host galaxies within the event localization region, their redshifts will contribute in a probabilistic way to the measurement of $H_0$, depending on the galaxies' distance and sky position. This approach
has been developed within a Bayesian framework by \citet{delpozzo} and \citet{chen17} and  {implemented} in \citet{fishbach} using GW170817{, which produced results consistent with the first measurement (\citealt{2017Natur.551...85A}) where the identified host galaxy, NGC 4993 (e.g., \citealt{palmese}), was used}. Eventually, a large sample of events will enable precise cosmological measurements using the dark siren approach. 

In this work, we measure $H_0$ using the gravitational wave event GW170814 (\citealt{gw170814}) as a dark siren. GW170814 resulted from the inspiral and merger of a binary black hole system at a {luminosity} distance of $540^{+130}_{-210}$ Mpc (median value with 90\% credible interval). The masses of the black holes were  $30.5^{+5.7}_{-3.0}$ and $25.3^{+2.8}_{-4.2} ~M_\odot$, each. GW170814 {is the first BBH detected by a triple network (including LIGO Hanford and Livingston, plus Virgo), and it} has the smallest localization volume of any of the BBH events detected by LVC thus far. {Therefore the number of potential host galaxies is lower compared to other events, making GW170814} the most appropriate event for this measurement. Additionally, the event localization region falls within the DES footprint, {making DES galaxy catalogs a prime sample for measurement of $H_0$}. With this one event, our goal is to provide a proof of principle measurement, addressing the challenges that are specific to the dark siren method, and establishing its potential to yield precision cosmology results in the near future. 

A key component of the measurement is crafting the appropriate galaxy catalog: completeness, as well as precise and accurate photometric redshifts (photo--$z$'s), throughout the entire volume probed are required. The overlap of GW170814's area with DES allows us to employ galaxy catalogs produced from the first three years of the survey (DES Y3; \citealt{dr1}). 
This first dark siren measurement is a step towards incorporating this new cosmological probe into the portfolio of cosmic surveys for dark energy. 

A detailed description of the data used in this analysis is provided in \S\ref{data}, followed by a description of our implementation of the method in \S\ref{method}. We present our results and discussion in \S\ref{results}, and our conclusions in \S\ref{conclusions}. Throughout this paper we assume a flat $\Lambda$CDM cosmology with $\Omega_m =0.3$ and $H_0$ values in the $20-140~{\rm km~s^{-1}~Mpc^{-1}}$ range. All quoted error bars represent the $68\%$ confidence level (CL), unless otherwise stated.

\section{Data}\label{data}

\begin{figure*}
\centering
\includegraphics[width=0.54\textwidth,trim=0 0 0 0, clip]{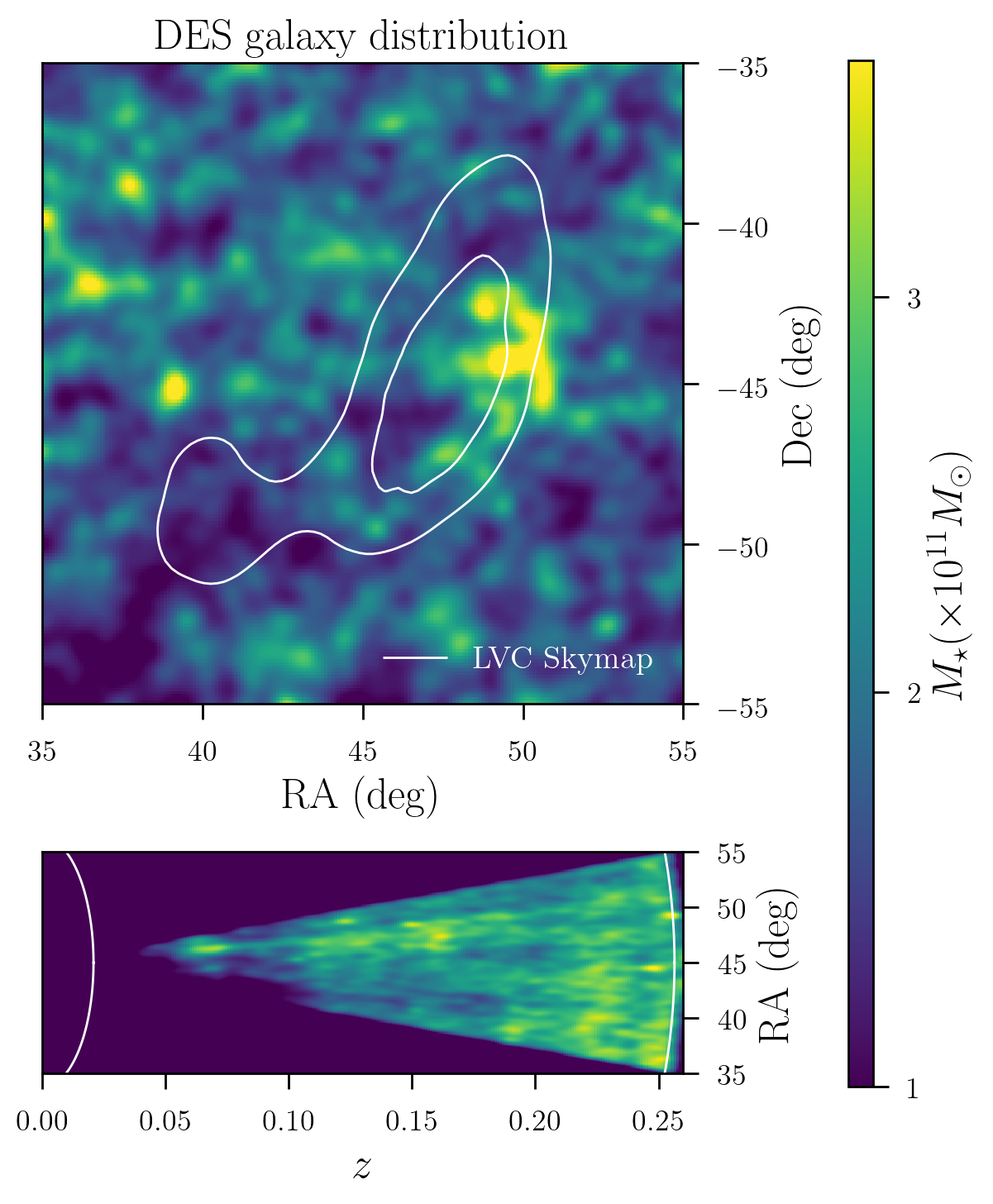}
\includegraphics[width=0.43\textwidth,trim=0 0 0.1 0.3, clip]{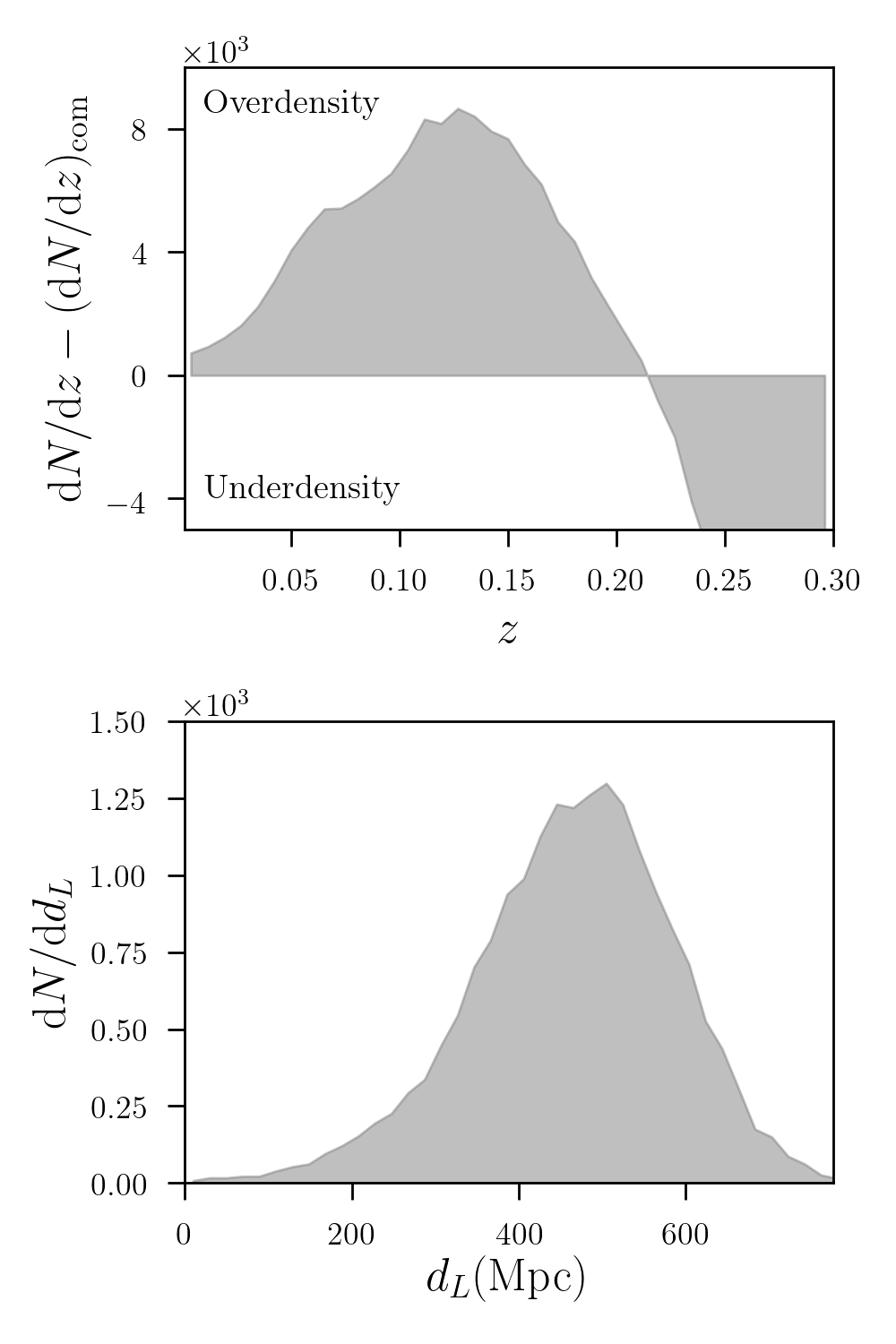}

\caption{\emph{Left:} Stellar mass distribution of the DES galaxies used in this analysis (color map) and the GW170814 localization region at 50 and 90\% CL (white contours). The region in redshift space is valid for the prior range $20<H_0<140$ $~{\rm km~s^{-1}~Mpc^{-1}}$. The stellar mass map has been smoothed with a Gaussian filter of width 0.3 deg.  {The bottom panel shows the galaxies' stellar mass distribution in RA and redshift, projected over the Dec.}
\emph{Right:} Distributions of the DES galaxy redshifts within the region of interest (top) and the luminosity distance in \textsc{HEALPix} pixels from the LVC distance likelihood, as given in the sky map (bottom). 
The histograms are obtained from a Monte--Carlo (MC) sampling the galaxies' redshift PDF and the luminosity distance likelihood in each pixel. The redshift distribution has been subtracted by a uniform distribution in comoving volume $(\de N/\de z)_{\rm com}$, {obtained assuming $H_0=70~{\rm km~s^{-1}~Mpc^{-1}}$}, and containing the same total number of galaxies to highlight the overdensity of galaxies in the region.}
\label{fig:data}
\end{figure*}

\subsection{The LVC sky map} 

The sky map used in this work is the publicly available \texttt{LALInference} map \citep{gcn_skymap}\footnote{\url{https://dcc.ligo.org/LIGO-T1700453/public}}, provided in \textsc{HEALPix} (\citealt{healpix}) pixels. 
The luminosity distance probability distribution is approximated with a Gaussian in each pixel. The region of interest, enclosing
90\% of the localization probability, is 61.66 deg$^2$. The projected sky map and the distribution of luminosity distance mean values {from the LVC distance likelihood} in each pixel within the region of interest are shown in Figure \ref{fig:data}.
The probability peak is located at RA, Dec = 
$(47.523, -44.856)$ deg. 
At the peak location, the luminosity distance is 504.7 Mpc and the Gaussian width is 91.9 Mpc. 
Using the limiting values of our $H_0$ prior range ([20,140] $ ~{\rm km~s^{-1}~Mpc^{-1}}$) we can convert the 90\% and $99.7\%$ distance range into a redshift range ($0.02<z<0.26$ and $z<0.3$, respectively) for this analysis.

\subsection{The DES galaxy catalog}

The DES\footnote{\url{www.darkenergysurvey.org}}(\citealt{descollaboration,2016MNRAS.460.1270D}) is an optical-near-infrared survey that images 5000 ${\rm deg}^2$ of the South Galactic Cap in the $grizY$ bands. The survey is being carried out using a $\sim3$ $\textrm{deg}^2$ CCD camera (the DECam, see \citealt{flaugher}) mounted on the Blanco 4-m telescope at the Cerro Tololo Inter-American Observatory (CTIO) in Chile. The data used here are from the first 3 years of observations (September 2013 -- February 2016, \citealt{dr1}). 
 
The DES Data Management (DESDM) pipeline was used for data reduction (\citealt{2018PASP..130g4501M}).
The process includes calibration of the single-epoch images, which are co--added after background subtraction and then cut into tiles. The source catalogue was created using \textsc{Source Extractor (SExtractor}, \citealt{sextractor}) to detect objects on the $riz$ co-added images. The median $10\sigma$ limiting magnitudes of Y3 data for galaxies are $g = 24.33$, $r = 24.08$, $i = 23.44$, $z = 22.69$, and $Y = 21.44$ mag (\citealt{dr1}). The photometry {used in this work is part of a value--added Y3 catalog not released with DR1, and} is the result of the Multi-Object Fitting (MOF) pipeline that uses the \texttt{ngmix} code.\footnote{\url{https://github.com/esheldon/ngmix}}
Following a procedure similar to \citet{firstyear} for Year 1 data, {the DES collaboration} made further selections to produce a high-quality object catalog called the Y3 ``gold'' catalog. For this sample, redshifts have been computed using the Directional Neighborhood Fitting (DNF; \citealt{dnf}), {and they are not included in DR1.}

The DNF method applied to Y3 data provides redshift information for each galaxy in the form of a probability distribution function (PDF), from which a mean redshift, and half of the central 68th percentile width are computed. 
The width of the PDF can be over or under-estimated due to the sampling of the training set and algorithmic details of DNF. This issue is particularly relevant for the redshift range used in this work, which is low compared to that exploited in weak lensing and large scale structure cosmology, for which the DNF method was optimized. We find that the typical uncertainty below redshift $z\sim 0.1$ is  underestimated by a factor of 10 when compared to the typical scatter found for the subset of the galaxies with available spectroscopic redshifts (where the standard deviation is $\sigma\gtrsim 0.015$). Thus, we add a minimum uncertainty of $0.015$ for these low--$z$ galaxies. At $0.1<z<0.3$, the uncertainty is well behaved and the average value follows $\bar{\sigma_z}(z) \simeq 0.013~ (1+z)^3$, as we find using an empirical fit.

We produce alternative photo--$z$ estimates with another machine learning code, \textsc{ANNz2} (\citealt{annz2}). This allows us to test the impact of the correction applied to the DNF errors on the posterior of the Hubble constant. {Photo--$z$ with \textsc{ANNz2} have previously been validated for cosmological analyses using DES Science Verification data (\citealt{bonnett,boris,2016PhRvD..94b2001A}) and for the Kilo--Degree Survey (KiDS; \citealt{bilicki}), and are produced as part of the DES photo--$z$ pipeline (\citealt{julia}).} In particular, it provides error estimates through a $k$--nearest neighbor ($k$NN) method, and dedicated redshifts for the purposes of this analysis. We additionally employ a reweighting technique (\citealt{lima}) specifically for our galaxy sample to further tune our redshifts. We run \textsc{ANNz2} in randomized regression mode with 50 Boosted Decision Trees (BDTs), using a spectroscopic sample of $245,458$ matching Y3 galaxies out to redshift $z\gtrsim 1$, randomly split into subsamples for training, testing and validation. The training and the reweighting use $griz$ MOF magnitudes. We find that the typical error roughly follows $\sim 0.02~(1+z)^3$ {in the redshift range of interest}. {The two algorithms, DNF and ANNz2, gave similar results, see section~\S\ref{results}}.

These redshifts, together with publicly available spectroscopic redshifts from 2dF, 6dF and SPT--GMOS \citep{2001MNRAS.328.1039C,2009MNRAS.399..683J,2016ApJS..227....3B} and the DES MOF photometry, are used to estimate galaxy properties (including stellar mass and absolute magnitude) of this sample. 
This is achieved through a broadband Spectral Energy Distribution (SED) fitting of galaxy magnitudes with \textsc{LePhare} (\citealt{arnouts}, \citealt{ilbertlephare}). Estimates of the galaxy properties used here from DES data alone have been tested and studied in several DES works (\citealt{palmese16,Etherington,palmese18}). We add a 0.05 systematic uncertainty in quadrature to the magnitudes, to account for systematic uncertainties in magnitude estimation and model variance.\footnote{This is a regularization to compensate for the synthetic model set grid and the fact that many SED fitting codes do not include a model error function. The value chosen is based on past experience of what gives stable results.} The simple stellar population (SSP) templates used {for the fitting} are \citet{bc03}, with three metallicities ($0.2~Z_\odot$, $Z_\odot$ and $2.5~Z_\odot$), a \citet{chabrier} Initial Mass Function (IMF) and a Milky Way \citep{allen} extinction law with five different values between 0 and 0.5 for the $E(B-V)$ reddening. The star formation history (SFH) chosen is exponentially declining as ${\rm e}^{-t/\tau}$ with $\tau=0.1,0.3,1,2,3,5,10,15$ and $30$ Gyr. 

The source list of the Y3 gold catalogue is $95\%$ complete for galaxies within our apparent magnitude limit, $r<23.35$ (\citealt{dr1}). This value is computed through the recovery rate of sources from the deeper CFHTLenS survey \citep{Erben13}, and thus includes the correct distribution of surface brightnesses. {Nevertheless, extended, low surface brightness galaxies near our flux limit may be preferentially missed by the detection pipeline. We therefore} provide an approximate completeness of sources throughout the redshift range of interest. Using DNF mean redshifts we convert the source completeness to $r<23.35$ from \citet{dr1} (Figure 12) into a completeness in redshift intervals, $\Delta z=0.02$. By taking the peak of the magnitude distribution in each bin as roughly our observed magnitude limit at that redshift, we find our sample is $>93\%$ complete across the range $0<z<0.26$. We further determined that the fraction of low redshift, extended galaxies missed by the DES Y3 pipeline is $\sim 1\%$, when compared with the 2MASS extended source catalog \citep{2012ApJS..199...26H}. For the purpose of this paper, we choose to ignore those ultra-low $z$ sources as most of them are at $z<0.02$ and are not relevant for the present analysis.

The DES Y3 gold catalog is nonetheless an observed magnitude--limited sample. This analysis requires a volume-limited sample, which we obtain by applying a luminosity cut. In order to determine the appropriate cut to create a volume--limited sample, we compute the completeness limits in terms of absolute quantities (luminositiy and stellar mass). We follow the method outlined in \cite{Pozzetti10} and \cite{Hartley13}. We identify galaxies with observed magnitudes that are bright enough to be complete and representative of the real galaxy population within redshift bins.
To compute the $95\%$ completeness limit in (rest--frame) luminosity, we scale the luminosities of this sample to that which they would have if their observed magnitude were equal to the survey completeness limit, and take the 95th percentile of the resulting luminosity distribution. 
This value corresponds to $-17.2$ in $r$-band absolute magnitude and $\sim 3.8 \times 10^8~{\rm M}_\odot$ in stellar mass for the redshift range of interest. We cut the DES catalog at the specific absolute luminosity value mentioned above. We conclude that our volume--limited galaxy sample is complete within the redshift range of interest for galaxies down to stellar masses of $\sim 3.8 \times 10^8~ {\rm M}_\odot$. In other words, our galaxy catalog contains $\sim 77\%$ of the total stellar mass in the volume considered by assuming that the galaxies follow a Schechter stellar mass function with the best fit values from \citet{weigel}. 

The final galaxy stellar mass and redshift distributions of galaxies are shown in Figure \ref{fig:data}. The stellar mass map clearly shows the presence of large scale structure, including clusters, voids and filaments. We recognize a number of well--known clusters within the volume of interest, including several Abell clusters. 
A uniform distribution of galaxies in comoving volume $(\de N/\de z)_{\rm com}$ has been subtracted from the observed galaxies' redshift distribution in Figure \ref{fig:data} to highlight the overdensities. The $(\de N/\de z)_{\rm com}$ distribution has been obtained by assuming $H_0=70~{\rm km~s^{-1}~Mpc^{-1}}$ and it contains the same total number of galaxies as the observed $\de N/\de z$ over the redshift range shown. We are able to identify a ``wall''--like structure around $z\sim0.06$ spanning most of the area between $35<{\rm RA}<55$ and $-55<{\rm Dec}<-35$, which is spectroscopically confirmed by 2dF, LCRS \citep{1996ApJ...470..172S}, and especially 6dF. A broader galaxy overdensity is found around $z\sim0.12$ (also seen in LCRS and 2dF, and composed of several Abell galaxy clusters). {This broad peak is also identified in redshift distributions by other photo--$z$ codes, including a template based code, the Bayesian Photometric Redshift (BPZ; \citealt{benitez})}.
We have further verified that overdensities at the lowest redshifts ($z\sim 0.06$) are also present in spectroscopic samples outside of the region of interest. This is expected at these low redshifts, where large scale structure projects onto vast areas of the sky. {In summary}, there are  $77,092$ 
galaxies within the $90\%$ {LIGO/Virgo} probability volume, and $105,011$ when $99.7\%$ of the distance probability is considered, of which $\sim 6,000$ have spectroscopic redshifts. 

\section{Method} \label{method}

In order to estimate the posterior probability of $H_0$ given GW data $d_{GW}$ from a single event detection, and electromagnetic (EM) data from a galaxy survey, we follow \citet{chen17}.
By applying Bayes' theorem, one can write the posterior as: 
\begin{equation}
p(H_0|d_{\rm GW},d_{\rm EM})  \propto p(d_{\rm GW},d_{\rm EM}|H_0)p(H_0)\, .
\label{eq:posterior}
\end{equation}
We assume that all cosmological parameters except for $H_0$ are fixed (Flat $\Lambda$CDM cosmology with $\Omega_m=0.3$ and $\Omega_\Lambda = 0.7$). 
We treat the joint GW and EM likelihood $p(d_{\rm GW},d_{\rm EM}|H_0)$ as the product of two individual likelihoods (since the processes involved in producing the data from the two experiments are independent) marginalized over all variables except for the true luminosity distance $d_L$ and solid angle 
$\hat{\Omega}_{\rm GW}$ of the GW source, and for the true host galaxy redshift $z_i$ and solid angle $\hat{\Omega}_i$. Note that the solid angles $\hat{\Omega}$ are vectors with the angular position of the source/galaxy as direction, and they all subtend the same area ($\sim 3\times 10^{-3}~{\rm deg^2}$) as the sky is pixelized with \textsc{HEALPix} maps in this work. If we assume that the event happened in one of the observed galaxies $i$, then $\hat{\Omega}_{\rm GW}$ and $\hat{\Omega}_i$ are related, and so are $d_L$ and $z_i$ through the cosmology (in this case, $H_0$). By marginalizing also over the choice of galaxy $i$, the joint, marginal likelihood can be written as:

\begin{equation}
\begin{split}
p(d_{\rm GW}, d_{\rm EM}|\{z_j,\hat{\Omega}_j\},H_0) \propto  \sum_i w_i \!\int\!\de d_L \, \de \hat{\Omega}_{\rm GW}\,p(d_{\rm GW}|d_L,\hat{\Omega}_{\rm GW}) \\
\times p(d_{\rm EM}|\{z_j,\hat{\Omega}_j\})\,\delta_D(d_L-d_L(z_i,H_0))
\,\delta_D(\hat{\Omega}_{\rm GW}-\hat{\Omega}_{i})\,,\label{eq:like}
\end{split}
\end{equation}
where $\delta_D$ is the Dirac delta function, {  $w_i$ are weights that represent the relative probability that different galaxies host a GW source, 
and {$\{z_j,\hat{\Omega}_j\}$} represents all the galaxies' redshift and solid angle.} These weights could be based on some galaxy properties, such as luminosity or star--formation rate, but here we assume they are uniform across all galaxies given our lack of knowledge of GW host galaxy properties.

We also need to marginalize over the galaxies' redshifts and sky positions, with a reasonable choice of prior $p(z_i,\Omega_i)$. If one assumes that the galaxies are uniformly distributed in comoving volume $V$, and volume--limited within $V_{\rm max}$:
\begin{equation}
\begin{split}
    p(z_i,\hat{\Omega}_i)~ \de z_i~ \de \hat{\Omega}_i & \propto \frac{1}{ V_{\rm max}} \frac{\de^2  V}{\de z_i \de \hat{\Omega}_i} \de z_i~ \de \hat{\Omega}_i \propto \frac{1}{V_{\rm max}} \frac{r^2(z_i)}{H(z_i)} \de z_i ~\de \hat{\Omega}_i\,,
    \end{split}
\end{equation}
where $r$ is the comoving distance to the galaxy. While this assumption holds on average over sufficiently large volumes, it is possible that future precision cosmology analyses will require taking into account the real clustering of galaxies in this formalism. 

{Assuming that we precisely know the galaxies' positions $\{\hat{\Omega}_j\}$ (which is realistic especially in the limit in which spatial probabilities are considered within \textsc{HEALPix} pixels), we can integrate over the galaxies' positions as delta functions about the observed values. The marginal EM likelihood reduces to $p(d_{\rm EM}|\{z_j\})$, which we approximate for simplicity by a product of Gaussian distributions, ${\mathcal N}$, for each galaxy, centred around the observed redshift values $z_{{\rm obs}, i}$ with a width given by the redshift's uncertainty $\sigma_{z,i}$ for each galaxy $i$: 
\begin{equation}
p(d_{\rm EM}|\{z_j\})= \prod_i p(z_{{\rm obs},i}|z_i)= \prod_i \mathcal{N}(z_{{\rm obs},i}, \sigma_{z,i}; z_i) .
\end{equation}
}

The marginal GW likelihood $p(d_{\rm GW}|d_L,\Omega)$ can be computed as prescribed in \citet{Singer}:
\begin{equation}
p(d_{\rm GW}|d_L,\hat{\Omega}) \propto p(\hat{\Omega}) \frac{1}{\sqrt{2\pi}\sigma(\hat{\Omega})}{\rm exp}\Big[ - \frac{\big(d_L-\mu (\hat{\Omega}) \big)^2}{2\sigma^2(\hat{\Omega})} \Big] N(\hat{\Omega})\,, \label{eq:dpdvmap} 
\end{equation}
where the position probability, location, normalization and scale (\texttt{PROB} $p(\hat{\Omega})$, \texttt{DISTMU} $\mu$, \texttt{DISTNORM} $N$ and \texttt{DISTSTD} $\sigma$ respectively) of the luminosity distance at each position are provided in the sky map. 

We now consider the selection effects of GW events and galaxies introduced by the experiments' sensitivities and detection pipelines. We follow the approach of \citet{chen17} and \citet{mandel}, and include a $[\beta(H_0)]^{-1}$ factor that normalizes the likelihood over all possible GW and EM data. Given that our galaxy catalog is volume--limited out to larger distances than the maximum observable distance for the GW events, this term reduces to: 
\begin{equation}
\beta(H_0) = \frac{V[d_{L,{\rm GW}}^{\rm max}(H_0)]}{V_{\rm max} (H_0)}\, ,
\end{equation}
where $V[d_{L,{\rm GW}}^{\rm max}(H_0)]$ is the maximum observable volume for the GW events considered.

Finally, Eq. (\ref{eq:posterior}) becomes:
\begin{widetext}
\begin{equation}
p(H_0|d_{\rm GW}, d_{\rm EM}) \propto \frac{p(H_0)}{V[d_{L,{\rm GW}}^{\rm max}(H_0)]} \sum_i \frac{1}{\mathcal{Z}_i}\int \de z_i \, p(d_{\rm GW}|d_L(z_i,H_0),\hat{\Omega}_i) p(d_{\rm EM}|z_i) \frac{r^2(z_i)}{H(z_i)} \, ,\label{eq:like2}
\end{equation}
\end{widetext}
where $\mathcal{Z}_i = \int p(d_{\rm EM} | z_i) r^2(z_i)/H(z_i) ~\de z_i$ {are evidence terms that arise from integrating out the other galaxy redshifts in each term of the sum.}
{This formalism can be extended to combine data $\{d_{{\rm GW},j}\}$ and $d_{{\rm EM}}$ from a sample of multiple events $j$, assuming that the GW events are independent and that the galaxy catalog is fixed for all events:
\begin{equation}
\begin{split}
p(H_0|\{d_{{\rm GW},j}\},d_{{\rm EM}}) 
\propto p(H_0) \int \de^N z_k \de^N \hat\Omega_k
p(z_k , \hat\Omega_k) \times \\
p(d_{EM} |  \{ z_k, \hat\Omega_k \} ) \left[ \prod_j p(d_{GW,j} | \{ z_k, \hat\Omega_k \} ) \right] \,.
\end{split}
\end{equation}
}

In the following, we assume a flat prior on $H_0$ within [20,140] $ ~{\rm km~s^{-1}~Mpc^{-1}}$, unless otherwise stated. This is a very broad prior, covering a range much larger than current estimates of $H_0$. {This choice was made as a compromise between the following aspects: i) a result which is mostly informed by the LVC and DES data rather than by external constraints, ii) a result which can be compared with the first standard siren estimate, and iii) a complete galaxy sample which contains most of the stellar mass within the localization volume, to minimize the chance of missing the real host galaxy. As explained in more detail in \S\ref{results}, the redshift cut is related to the $H_0$ prior range, and in order to explore higher values of $H_0$, one needs to include higher redshift galaxies, and make a higher luminosity cut to preserve the volume limited sample.}

A blinded analysis has been performed when estimating the $H_0$ posterior from the data to avoid confirmation bias. The values of the Hubble constant have been randomly displaced by an unknown amount, and we unblinded after our pipeline was able to reliably reproduce the input cosmology on simulation tests.

\section{Results and Discussion} \label{results}
\begin{figure*}
    \centering
    \includegraphics[width=0.8\linewidth]{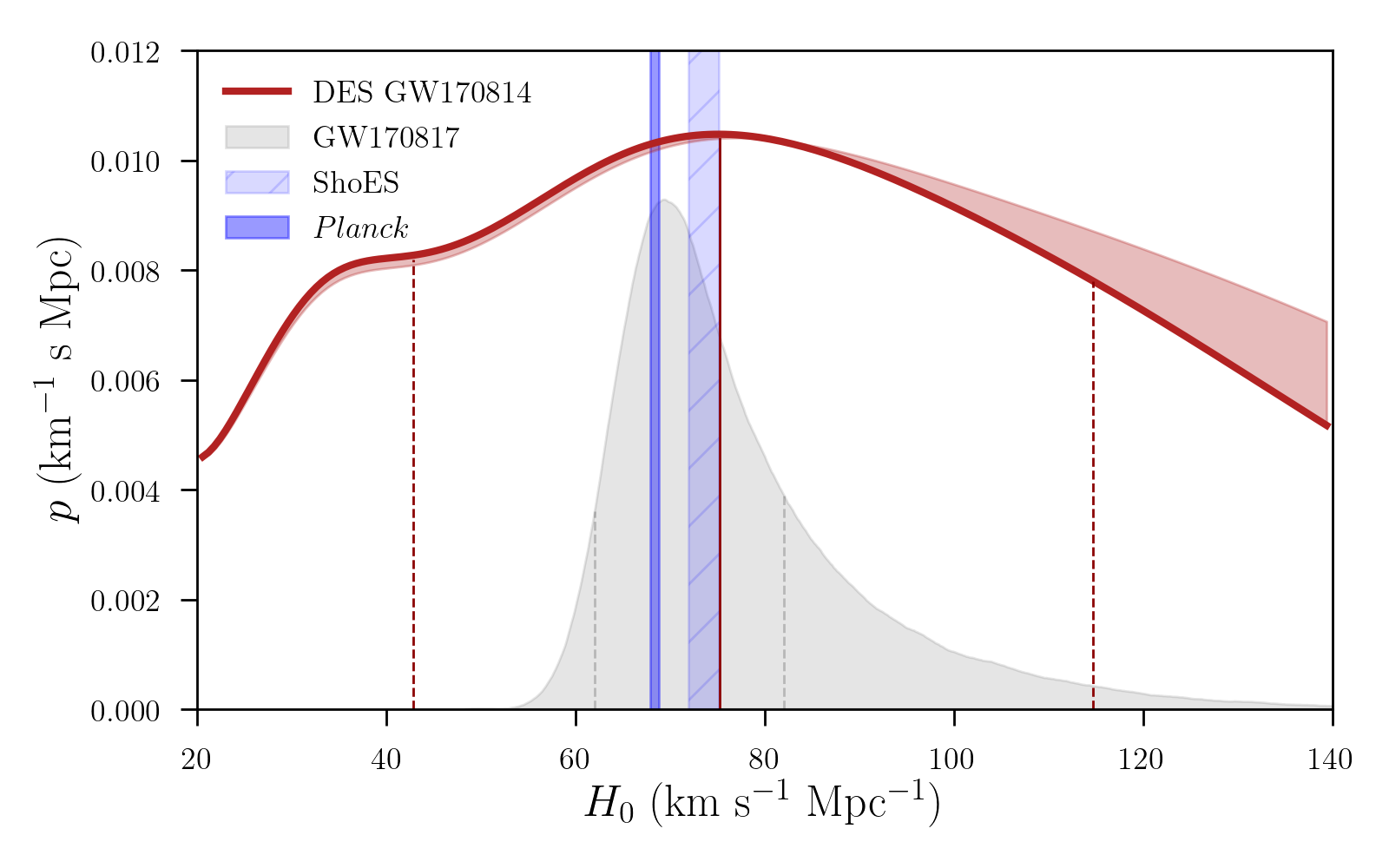}
    \caption{{Hubble constant posterior distribution obtained by marginalizing over $\sim 77,000$ possible host galaxies (red line), showing the maximum value (solid vertical line). The maximum a posteriori and its 68\% confidence level is $H_0 = 75^{+40}_{- 32}~{\rm km~s^{-1}~Mpc^{-1}}$ for a flat prior in the range [20,140] ${\rm km~s^{-1}~Mpc^{-1}}$. The shaded region represents the change in the posterior when different fractions of the localization volume are considered (from 90 to 99.7\% of the LIGO/Virgo luminosity distance posterior). The PDF computed from the larger volume has been renormalized to have the same value of the 90\% localization volume $H_0$ posterior at the maximum, to highlight differences below and beyond the main peak. The posterior obtained by \citet{2017Natur.551...85A} for the bright standard siren event GW170817, associated to one galaxy, is shown in grey. The prior used in that work was flat--in--log over a narrower range ([50,140] ${\rm km~s^{-1}~Mpc^{-1}}$), and the posterior has been rescaled by a factor 0.2 for visualization purposes. The 68\% CL of both PDFs is shown by the dashed lines.}
    Constraints from \emph{Planck} (\citealt{planck18}) and SHoES (\citealt{shoes,riess18}) at 1$\sigma$ are shown in purple boxes.}
    \label{fig:result}
\end{figure*}

We apply the described methodology to the DES galaxies' redshifts and the GW170814 LIGO/Virgo sky map to produce a posterior distribution for the Hubble constant. We find that changes in the $H_0$ estimate and its uncertainty between using the corrected DNF photo--$z$'s or the \textsc{ANNz2} outputs are below the percent level. 
This agreement is expected, since the two methods produce  redshift distributions that are consistent with similar uncertainties. 
We also add a $0.001$ systematic {redshift} error in quadrature (corresponding to a typical peculiar velocity of $\sim 300~{\rm km~s^{-1}}$). The effect of this correction on the posterior is negligible because only a few percent of the galaxies have a spectroscopic redshift, and the effect of peculiar velocities on the remaining galaxies is more than an order of magnitude below their typical photo--$z$ error. 

Our maximum a posteriori estimate of the Hubble constant is $H_0 = 75^{+40}_{- 32}~{\rm km~s^{-1}~Mpc^{-1}}$ using a flat prior between 20 and 140 ${\rm km~s^{-1}~Mpc^{-1}}$. The full posterior distribution is shown in Figure \ref{fig:result}, and Table \ref{tab:results} summarizes our findings. The presence of a main, though broad, peak, is expected given the large scale structure seen in the observed volume.

{As described in section 2.1, the galaxy sample used in these results is selected as described in \S\ref{data}, and 
{covers the LIGO/Virgo 90\% credible localization volume.} The distance cut is  translated into a redshift cut (made on the mean photo-$z$ value of each galaxy) for a given $H_0$ prior. This cut ensures that the galaxy catalog is as complete as possible throughout the whole redshift range of interest for the cosmological parameters used, and includes the fainter galaxies observable for a volume--limited sample defined as in \S\ref{data}. In fact, in order to include more distant galaxies, the luminosity cut needs to be brighter to ensure that the sample is still volume--limited, with the risk of missing the true host galaxy. We have explored the impact of the redshift cut on the $H_0$ posterior, while keeping the angular selection to be within the 90\% credible localization area. The effect of including galaxies out to $99.7\%$ of the distance localization (corresponding to $z\lesssim 0.3$) is most pronounced at high $H_0$ values, as shown by the shaded red region in Figure \ref{fig:result}. With this less restrictive cut, the credible region shifts to $H_0 = 77^{+41}_{- 33}~{\rm km~s^{-1}~Mpc^{-1}}$, showing a $\sim 2\%$ change of the maximum.  
The effect described here arises from tens of thousand of galaxies at the higher redshifts included with the more relaxed distance cut and the ansatz of Gaussianity of the luminosity distance posterior. 
In fact, these galaxies contribute with a non--negligible probability to the posterior because of the high $d_L$ tail shown in the bottom right panel of Figure \ref{fig:data}, and they contribute more significantly at high $H_0$ values. This few percent effect is insignificant at the current levels of precision, but will need to be explored in the future using a more realistic luminosity distance posterior.}

Our result agrees well (as expected, due to the large uncertainty)  with the latest CMB estimate of the Hubble constant by the \emph{Planck} Collaboration ($67.36 \pm 0.54~{\rm km~s^{-1}~Mpc^{-1}}$ from TT,TE,EE+lowP+lensing; \citealt{planck18}), {and with results using distance ladder methods} by ShoES ($73.52 \pm 1.62~{\rm km~s^{-1}~Mpc^{-1}}$; \citealt{shoes}) and by DES ($67.77 \pm 1.30 ~{\rm km~s^{-1}~Mpc^{-1}}$ from SN+BAO; \citealt{2018arXiv181102376M}).

For the bright standard siren measurement using GW170817 and its electromagnetic counterpart, \citet{2017Natur.551...85A} found $H_0 = 70.0^{+ 12.0}_{-8.0}~{\rm km~s^{-1}~Mpc^{-1}}$ at 68\% credible interval. Without an EM counterpart leading to a unique host galaxy redshift, we would have recovered a broader $H_0$ posterior since we average over all possible host galaxies in the localization volume. For example, \citet{fishbach} applied the statistical standard siren method to GW170817 and found a larger uncertainty than the counterpart standard siren result: $H_0 = 76^{+ 48}_{-23}~{\rm km~s^{-1}~Mpc^{-1}}$ for a uniform prior over the range $[10,220]~{\rm km~s^{-1}~Mpc^{-1}}$. For a BBH standard siren measurement, as in this work, the combination of the larger localization volume (implying a significantly greater number of potential host galaxies) and the large photometric redshift uncertainty for each galaxy results in an even broader $H_0$ posterior. Therefore, while applying the statistical standard siren method to GW170817 yields a 68\% credible region on $H_0$ comprising 34\% of the prior range \citep{fishbach}, in this work we obtain a 68\% credible region on $H_0$ that is 60\% of the prior range. We note that the prior used in \citet{fishbach} is 1.75 times broader than the prior used in this work; if we adopt the same broader prior of [10,220] for our analysis of GW170814, we find $H_0 = 78^{+ 96}_{-24}~{\rm km~s^{-1}~Mpc^{-1}}$. {The dependence of the width of the $H_0$ posterior on the prior width is a consequence of the fact that the GW observation, which provides only a luminosity distance estimate, is consistent with arbitrarily large $H_0$'s, if there are galaxies at sufficiently large redshifts. If the galaxy catalogue extends to some redshift, $z_{\rm max}$, the posterior would fall off around $H_0 \approx c z_{\rm max}/d_L$, where $d_L$ is the typical luminosity distance from the GW posterior. However, this fall off is artificial since there are galaxies at greater redshifts which are not included in the catalogue. These may be accounted for using catalogue incompleteness corrections. We chose the prior range for this analysis rather than a larger one such that we did not need to include such corrections, which simplifies the analysis. However, dark siren measurements will become particularly interesting when multiple events can be combined and this effect becomes irrelevant (\citealt{chen17}).}

The analysis in \citet{fishbach} for GW170817 used the GLADE galaxy catalog (\citealt{glade}), and accounted for incompleteness at the distance of GW170817. GLADE becomes significantly incomplete at the distance to GW170814. As GW detectors improve in sensitivity, the majority of dark standard sirens will be detected at even greater distances and with larger localization volumes, well beyond the reach of spectroscopic galaxy catalogs. This highlights the need for reliable and complete photometric galaxy catalogs. Surveys such as DES, Pan--STARRS1 (\citealt{panstarrs}) and LSST are therefore likely to play an important role in future constraints from BBH standard sirens.

\begin{table}[]
\centering
\begin{tabular}{cccccc}
 Prior  & $H_0 $ & $+\sigma_{H_0}$ & $-\sigma_{H_0}$ & $\sigma_{H_0}/H_0$ & $\sigma_{H_0}/H_0$ prior\\
\hline
 $[20,140]$ & 75 & 40 & 32 & 47.8\% & 54.3\%\\
\end{tabular}
\caption{Hubble constant estimate from this work. All $H_0$ values and errors are in $~{\rm km~s^{-1}~Mpc^{-1}}$. {The uncertainty from the flat prior only is derived by assuming the same $H_0$ maximum found in the analysis.} Quoted uncertainties represent 68\% confidence level around the maximum of the posterior, {and they are statistical only}. {The last column quantity 
($\sigma_{H_0}/H_0$ prior) corresponds to 68\% times the 
prior width divided by $H_0$.}}
\label{tab:results}
\end{table}

The assumption throughout this work is that even if the event occurred in a galaxy below our luminosity threshold, large scale structure predicts that fainter galaxies follow the clustering pattern of the more luminous galaxies in our sample. 
We have verified in our simulations that a threshold up to 1 magnitude brighter than the limit used here to place events has a negligible impact over a sample of 100 events, provided that the catalog is volume--limited for the range of redshifts relevant to the measurement. 

Since galaxies are biased tracers of the Universe's dark matter, some theories predict that the origin of the black holes involved in these GW events is primordial, constituting part or all of the dark matter (\citealt{PhysRevLett.116.201301,2017PDU....15..142C,pbh,pbh2}). In that case, GW events follow exactly the underlying dark matter distribution (presenting an unbiased tracer). Because of the stellar mass to dark matter halo connection (see \citealt{2018ARA&A..56..435W} and references therein) it is reasonable to weight galaxies by their stellar mass in Eq. (\ref{eq:like}) as $w_i\propto M_\star$. {The impact of this scaling with stellar mass or star--formation rate has been explored in \citet{fishbach}.} We find that the stellar mass weighting has a negligible effect on the posterior. This is due to the large volume analyzed (over which the stellar masses tend to be averaged out) and to the precision level of this measurement. In other theories, these black hole binaries are produced in very low metallicity galaxies (e.g. \citealt{2018MNRAS.474.4997C,2018MNRAS.481.5324M}), biased relative to the dark matter distribution differently than the luminous galaxies in our catalog. Annis et al. (in prep.) explore the effect of the tracer bias assumptions on the $H_0$ posterior for future analyses aiming at precision measurements.

Another assumption of our analysis that needs attention concerns the redshift likelihood. As anticipated above for the GW likelihood, this will not, in general, be well approximated by a Gaussian. In the future, we plan to explore the impact of realistic photometric redshift PDFs on the $H_0$ posterior, in order to enable precision cosmology with binary--black--hole events. An analysis with the full, asymmetric, GW likelihood will also be required. 
While an estimate of those effects is needed, tests on off--source lines--of--sight show that our constraint is likely not strongly impacted by the photo--$z$ training sample or systematic failures.

In the past two LVC observing seasons, black--hole mergers outnumbered neutron star events at a rate of approximately 10 to 1. Uncertainties on the expected detection rate are large, but conservative estimates predict $\sim 1$ event per week for the upcoming observing campaign (scheduled to start in April 2019). The majority of these events will have larger localization volumes than GW170814 (\citealt{chen12} estimate $\lesssim 1\%$ of BBHs will be localized to better than $10^4~$Mpc$^3$) and hence provide poorer constraints than those reported here. However, given the high expected event rate for dark sirens, larger event samples will be available in the future. {\citet{chen17} provide forecasts using a distribution of realistic localization area, finding that the dark siren method will reach $\sim10\%$ statistical precision on $H_0$ by 2026 from BBH mergers only, and 5--10\% precision from BNS mergers if none of them have EM counterparts.}

\section{Conclusions}\label{conclusions}

In this paper, we have performed the first measurement of the Hubble constant using a gravitational wave detection of a binary--black--hole merger as a dark standard siren and the DES galaxies as a sample of potential host galaxies. Our analysis was blinded to avoid confirmation bias. Our main results, discussed in \S\ref{results}, include a measurement of $H_0 = 75^{+40}_{- 32}~{\rm km~s^{-1}~Mpc^{-1}}$ for a flat prior within [20,140] km s$^{-1}$ Mpc$^{-1}$, consistent with previous measurements of $H_0$. The $68\%$ confidence interval quoted here is $60\%$ of the uniform prior range, {and it depends on the width of the prior. For example, with a broader prior of [10,220] ${\rm km~s^{-1}~Mpc^{-1}}$, we find $H_0 = 78^{+ 96}_{-24}~{\rm km~s^{-1}~Mpc^{-1}}$. Albeit weak, this measurement is not uninformative and the method becomes more powerful when we combine large numbers of dark sirens~\citep{chen17}. }

{Future dark siren measurements will require complete galaxy catalogs. A wide field galaxy catalog with a DES--like depth is currently available only for $\sim 1/8$ of the sky. However, DES can be complemented with other datasets taken with DECam (such as the Blanco Imaging of the Southern Sky, BLISS, and the Dark Energy Camera Legacy Survey, DECals), to cover the whole Southern sky to a good depth ($r\sim 23.4$, 5$\sigma$ depth).} An even deeper survey with more precise photo--z's, such as the Large Synoptic Survey Telescope (LSST; \citealt{2012arXiv1211.0310L}), would be of great value for further improving these constraints.

At the expected level of precision from hundreds of events ($<10\%$), systematics will play an important role. In future work, we plan to incorporate systematic uncertainties in our simulated data studies, in order to prepare for precision cosmology analyses on real data. We anticipate that some of the main sources of systematics will be photo--$z$ biases and catastrophic outliers, photo--$z$ training sample sample variance, galaxy catalog cuts and galaxy catalog completeness. 
In order to achieve the full potential of statistical standard siren cosmology, wide and deep galaxy surveys such as DES and LSST are necessary. Overall, our findings show that the synergy between gravitational wave black--hole merger detections and new generation large galaxy surveys will establish a new powerful probe for precision cosmology. 

\acknowledgments
Funding for the DES Projects has been provided by the DOE and NSF(USA), MEC/MICINN/MINECO(Spain), STFC(UK), HEFCE(UK). NCSA(UIUC), KICP(U. Chicago), CCAPP(Ohio State), 
MIFPA(Texas A\&M), CNPQ, FAPERJ, FINEP (Brazil), DFG(Germany) and the Collaborating Institutions in the Dark Energy Survey.

The Collaborating Institutions are Argonne Lab, UC Santa Cruz, University of Cambridge, CIEMAT-Madrid, University of Chicago, University College London, 
DES-Brazil Consortium, University of Edinburgh, ETH Z{\"u}rich, Fermilab, University of Illinois, ICE (IEEC-CSIC), IFAE Barcelona, Lawrence Berkeley Lab, 
LMU M{\"u}nchen and the associated Excellence Cluster Universe, University of Michigan, NOAO, University of Nottingham, Ohio State University, University of 
Pennsylvania, University of Portsmouth, SLAC National Lab, Stanford University, University of Sussex, Texas A\&M University, and the OzDES Membership Consortium.

Based in part on observations at Cerro Tololo Inter-American Observatory, National Optical Astronomy Observatory, which is operated by the Association of 
Universities for Research in Astronomy (AURA) under a cooperative agreement with the National Science Foundation.

The DES Data Management System is supported by the NSF under Grant Numbers AST-1138766 and AST-1536171. The DES participants from Spanish institutions are partially 
supported by MINECO under grants AYA2015-71825, ESP2015-88861, FPA2015-68048, and Centro de Excelencia SEV-2016-0588, SEV-2016-0597 and MDM-2015-0509. Research leading 
to these results has received funding from the ERC under the EU's 7$^{\rm th}$ Framework Programme including grants ERC 240672, 291329 and 306478.
We acknowledge support from the Australian Research Council Centre of Excellence for All-sky Astrophysics (CAASTRO), through project number CE110001020.

This manuscript has been authored by Fermi Research Alliance, LLC under Contract No. DE-AC02-07CH11359 with the U.S. Department of Energy, Office of Science, Office of High Energy Physics. The United States Government retains and the publisher, by accepting the article for publication, acknowledges that the United States Government retains a non-exclusive, paid-up, irrevocable, world-wide license to publish or reproduce the published form of this manuscript, or allow others to do so, for United States Government purposes.

\vspace{\baselineskip}
The authors gratefully acknowledge the support of the United States
National Science Foundation (NSF) for the construction and operation of the
LIGO Laboratory and Advanced LIGO as well as the Science and Technology Facilities Council (STFC) of the
United Kingdom, the Max-Planck-Society (MPS), and the State of
Niedersachsen/Germany for support of the construction of Advanced LIGO 
and construction and operation of the GEO600 detector. 
Additional support for Advanced LIGO was provided by the Australian Research Council.
The authors gratefully acknowledge the Italian Istituto Nazionale di Fisica Nucleare (INFN),  
the French Centre National de la Recherche Scientifique (CNRS) and
the Foundation for Fundamental Research on Matter supported by the Netherlands Organisation for Scientific Research, 
for the construction and operation of the Virgo detector
and the creation and support  of the EGO consortium. 
The authors also gratefully acknowledge research support from these agencies as well as by 
the Council of Scientific and Industrial Research of India, 
the Department of Science and Technology, India,
the Science \& Engineering Research Board (SERB), India,
the Ministry of Human Resource Development, India,
the Spanish  Agencia Estatal de Investigaci\'on,
the Vicepresid\`encia i Conselleria d'Innovaci\'o, Recerca i Turisme and the Conselleria d'Educaci\'o i Universitat del Govern de les Illes Balears,
the Conselleria d'Educaci\'o, Investigaci\'o, Cultura i Esport de la Generalitat Valenciana,
the National Science Centre of Poland,
the Swiss National Science Foundation (SNSF),
the Russian Foundation for Basic Research, 
the Russian Science Foundation,
the European Commission,
the European Regional Development Funds (ERDF),
the Royal Society, 
the Scottish Funding Council, 
the Scottish Universities Physics Alliance, 
the Hungarian Scientific Research Fund (OTKA),
the Lyon Institute of Origins (LIO),
the Paris \^{I}le-de-France Region, 
the National Research, Development and Innovation Office Hungary (NKFIH), 
the National Research Foundation of Korea,
Industry Canada and the Province of Ontario through the Ministry of Economic Development and Innovation, 
the Natural Science and Engineering Research Council Canada,
the Canadian Institute for Advanced Research,
the Brazilian Ministry of Science, Technology, Innovations, and Communications,
the International Center for Theoretical Physics South American Institute for Fundamental Research (ICTP-SAIFR), 
the Research Grants Council of Hong Kong,
the National Natural Science Foundation of China (NSFC),
the Leverhulme Trust, 
the Research Corporation, 
the Ministry of Science and Technology (MOST), Taiwan
and
the Kavli Foundation.
The authors gratefully acknowledge the support of the NSF, STFC, MPS, INFN, CNRS and the
State of Niedersachsen/Germany for provision of computational resources.


\bibliographystyle{yahapj}
\bibliography{references}

\begin{thebibliography}{}
\providecommand\natexlab[1]{#1}
\providecommand\JournalTitle[1]{#1}

\bibitem[{Abbott {et~al.}(2016)Abbott, Abbott, Abbott, Abernathy, Acernese,
  Ackley, Adams, Adams, Addesso, Adhikari, Adya, Affeldt, Agathos, Agatsuma,
  Aggarwal, Aguiar, Aiello, Ain, Ajith, Allen, Allocca, Altin, Anderson,
  Anderson, Arai, Arain, Araya, Arceneaux, Areeda, Arnaud, Arun, Ascenzi,
  Ashton, Ast, Aston, Astone, Aufmuth, Aulbert, Babak, Bacon, Bader, Baker,
  Baldaccini, Ballardin, Ballmer, Barayoga, Barclay, Barish, Barker, Barone,
  Barr, Barsotti, Barsuglia, Barta, Bartlett, Barton, Bartos, Bassiri, Basti,
  Batch, Baune, Bavigadda, Bazzan, Behnke, Bejger, Belczynski, Bell, Bell,
  Berger, Bergman, Bergmann, Berry, Bersanetti, Bertolini, Betzwieser, Bhagwat,
  Bhandare, Bilenko, Billingsley, Birch, Birney, Birnholtz, Biscans, Bisht,
  Bitossi, Biwer, Bizouard, Blackburn, Blair, Blair, Blair, Bloemen, Bock,
  Bodiya, Boer, Bogaert, Bogan, Bohe, Bojtos, Bond, Bondu, Bonnand, Boom, Bork,
  Boschi, Bose, Bouffanais, Bozzi, Bradaschia, Brady, Braginsky, Branchesi,
  Brau, Briant, Brillet, Brinkmann, Brisson, Brockill, Brooks, Brown, Brown,
  Brown, Buchanan, Buikema, Bulik, Bulten, Buonanno, Buskulic, Buy, Byer,
  Cabero, Cadonati, Cagnoli, Cahillane, Bustillo, Callister, Calloni, Camp,
  Cannon, Cao, Capano, Capocasa, Carbognani, Caride, Diaz, Casentini, Caudill,
  Cavagli\`a, Cavalier, Cavalieri, Cella, Cepeda, Baiardi, Cerretani, Cesarini,
  Chakraborty, Chalermsongsak, Chamberlin, Chan, Chao, Charlton,
  Chassande-Mottin, Chen, Chen, Cheng, Chincarini, Chiummo, Cho, Cho, Chow,
  Christensen, Chu, Chua, Chung, Ciani, Clara, Clark, Cleva, Coccia, Cohadon,
  Colla, Collette, Cominsky, Constancio, Conte, Conti, Cook, Corbitt, Cornish,
  Corsi, Cortese, Costa, Coughlin, Coughlin, Coulon, Countryman, Couvares,
  Cowan, Coward, Cowart, Coyne, Coyne, Craig, Creighton, Creighton, Cripe,
  Crowder, Cruise, Cumming, Cunningham, Cuoco, Canton, Danilishin, D'Antonio,
  Danzmann, Darman, Da~Silva~Costa, Dattilo, Dave, Daveloza, Davier, Davies,
  Daw, Day, De, DeBra, Debreczeni, Degallaix, De~Laurentis, Del\'eglise,
  Del~Pozzo, Denker, Dent, Dereli, Dergachev, DeRosa, De~Rosa, DeSalvo,
  Dhurandhar, D\'{\i}az, Di~Fiore, Di~Giovanni, Di~Lieto, Di~Pace, Di~Palma,
  Di~Virgilio, Dojcinoski, Dolique, Donovan, Dooley, Doravari, Douglas, Downes,
  Drago, Drever, Driggers, Du, Ducrot, Dwyer, Edo, Edwards, Effler, Eggenstein,
  Ehrens, Eichholz, Eikenberry, Engels, Essick, Etzel, Evans, Evans, Everett,
  Factourovich, Fafone, Fair, Fairhurst, Fan, Fang, Farinon, Farr, Farr,
  Favata, Fays, Fehrmann, Fejer, Feldbaum, Ferrante, Ferreira, Ferrini,
  Fidecaro, Finn, Fiori, Fiorucci, Fisher, Flaminio, Fletcher, Fong, Fournier,
  Franco, Frasca, Frasconi, Frede, Frei, Freise, Frey, Frey, Fricke, Fritschel,
  Frolov, Fulda, Fyffe, Gabbard, Gair, Gammaitoni, Gaonkar, Garufi, Gatto,
  Gaur, Gehrels, Gemme, Gendre, Genin, Gennai, George, Gergely, Germain, Ghosh,
  Ghosh, Ghosh, Giaime, Giardina, Giazotto, Gill, Glaefke, Gleason, Goetz,
  Goetz, Gondan, Gonz\'alez, Castro, Gopakumar, Gordon, Gorodetsky, Gossan,
  Gosselin, Gouaty, Graef, Graff, Granata, Grant, Gras, Gray, Greco, Green,
  Greenhalgh, Groot, Grote, Grunewald, Guidi, Guo, Gupta, Gupta, Gushwa,
  Gustafson, Gustafson, Hacker, Hall, Hall, Hammond, Haney, Hanke, Hanks,
  Hanna, Hannam, Hanson, Hardwick, Harms, Harry, Harry, Hart, Hartman, Haster,
  Haughian, Healy, Heefner, Heidmann, Heintze, Heinzel, Heitmann, Hello,
  Hemming, Hendry, Heng, Hennig, Heptonstall, Heurs, Hild, Hoak, Hodge, Hofman,
  Hollitt, Holt, Holz, Hopkins, Hosken, Hough, Houston, Howell, Hu, Huang,
  Huerta, Huet, Hughey, Husa, Huttner, Huynh-Dinh, Idrisy, Indik, Ingram, Inta,
  Isa, Isac, Isi, Islas, Isogai, Iyer, Izumi, Jacobson, Jacqmin, Jang, Jani,
  Jaranowski, Jawahar, Jim\'enez-Forteza, Johnson, Johnson-McDaniel, Jones,
  Jones, Jonker, Ju, Haris, Kalaghatgi, Kalogera, Kandhasamy, Kang, Kanner,
  Karki, Kasprzack, Katsavounidis, Katzman, Kaufer, Kaur, Kawabe, Kawazoe,
  K\'ef\'elian, Kehl, Keitel, Kelley, Kells, Kennedy, Keppel, Key,
  Khalaidovski, Khalili, Khan, Khan, Khan, Khazanov, Kijbunchoo, Kim, Kim, Kim,
  Kim, Kim, Kim, King, King, Kinzel, Kissel, Kleybolte, Klimenko, Koehlenbeck,
  Kokeyama, Koley, Kondrashov, Kontos, Koranda, Korobko, Korth, Kowalska,
  Kozak, Kringel, Krishnan, Kr\'olak, Krueger, Kuehn, Kumar, Kumar, Kuo,
  Kutynia, Kwee, Lackey, Landry, Lange, Lantz, Lasky, Lazzarini, Lazzaro,
  Leaci, Leavey, Lebigot, Lee, Lee, Lee, Lee, Lenon, Leonardi, Leong, Leroy,
  Letendre, Levin, Levine, Li, Libson, Littenberg, Lockerbie, Logue, Lombardi,
  London, Lord, Lorenzini, Loriette, Lormand, Losurdo, Lough, Lousto, Lovelace,
  L\"uck, Lundgren, Luo, Lynch, Ma, MacDonald, Machenschalk, MacInnis, Macleod,
  Maga\~na Sandoval, Magee, Mageswaran, Majorana, Maksimovic, Malvezzi, Man,
  Mandel, Mandic, Mangano, Mansell, Manske, Mantovani, Marchesoni, Marion,
  M\'arka, M\'arka, Markosyan, Maros, Martelli, Martellini, Martin, Martin,
  Martynov, Marx, Mason, Masserot, Massinger, Masso-Reid, Matichard, Matone,
  Mavalvala, Mazumder, Mazzolo, McCarthy, McClelland, McCormick, McGuire,
  McIntyre, McIver, McManus, McWilliams, Meacher, Meadors, Meidam, Melatos,
  Mendell, Mendoza-Gandara, Mercer, Merilh, Merzougui, Meshkov, Messenger,
  Messick, Meyers, Mezzani, Miao, Michel, Middleton, Mikhailov, Milano, Miller,
  Millhouse, Minenkov, Ming, Mirshekari, Mishra, Mitra, Mitrofanov,
  Mitselmakher, Mittleman, Moggi, Mohan, Mohapatra, Montani, Moore, Moore,
  Moraru, Moreno, Morriss, Mossavi, Mours, Mow-Lowry, Mueller, Mueller, Muir,
  Mukherjee, Mukherjee, Mukherjee, Mukund, Mullavey, Munch, Murphy, Murray,
  Mytidis, Nardecchia, Naticchioni, Nayak, Necula, Nedkova, Nelemans, Neri,
  Neunzert, Newton, Nguyen, Nielsen, Nissanke, Nitz, Nocera, Nolting,
  Normandin, Nuttall, Oberling, Ochsner, O'Dell, Oelker, Ogin, Oh, Oh, Ohme,
  Oliver, Oppermann, Oram, O'Reilly, O'Shaughnessy, Ott, Ottaway, Ottens,
  Overmier, Owen, Pai, Pai, Palamos, Palashov, Palomba, Pal-Singh, Pan, Pan,
  Pankow, Pannarale, Pant, Paoletti, Paoli, Papa, Paris, Parker, Pascucci,
  Pasqualetti, Passaquieti, Passuello, Patricelli, Patrick, Pearlstone,
  Pedraza, Pedurand, Pekowsky, Pele, Penn, Perreca, Pfeiffer, Phelps, Piccinni,
  Pichot, Pickenpack, Piergiovanni, Pierro, Pillant, Pinard, Pinto, Pitkin,
  Poeld, Poggiani, Popolizio, Post, Powell, Prasad, Predoi, Premachandra,
  Prestegard, Price, Prijatelj, Principe, Privitera, Prix, Prodi, Prokhorov,
  Puncken, Punturo, Puppo, P\"urrer, Qi, Qin, Quetschke, Quintero,
  Quitzow-James, Raab, Rabeling, Radkins, Raffai, Raja, Rakhmanov, Ramet,
  Rapagnani, Raymond, Razzano, Re, Read, Reed, Regimbau, Rei, Reid, Reitze,
  Rew, Reyes, Ricci, Riles, Robertson, Robie, Robinet, Rocchi, Rolland,
  Rollins, Roma, Romano, Romano, Romanov, Romie, Rosi\ifmmode~\acute{n}\else
  \'{n}\fi{}ska, Rowan, R\"udiger, Ruggi, Ryan, Sachdev, Sadecki, Sadeghian,
  Salconi, Saleem, Salemi, Samajdar, Sammut, Sampson, Sanchez, Sandberg,
  Sandeen, Sanders, Sanders, Sassolas, Sathyaprakash, Saulson, Sauter, Savage,
  Sawadsky, Schale, Schilling, Schmidt, Schmidt, Schnabel, Schofield,
  Sch\"onbeck, Schreiber, Schuette, Schutz, Scott, Scott, Sellers, Sengupta,
  Sentenac, Sequino, Sergeev, Serna, Setyawati, Sevigny, Shaddock, Shaffer,
  Shah, Shahriar, Shaltev, Shao, Shapiro, Shawhan, Sheperd, Shoemaker,
  Shoemaker, Siellez, Siemens, Sigg, Silva, Simakov, Singer, Singer, Singh,
  Singh, Singhal, Sintes, Slagmolen, Smith, Smith, Smith, Smith, Son, Sorazu,
  Sorrentino, Souradeep, Srivastava, Staley, Steinke, Steinlechner,
  Steinlechner, Steinmeyer, Stephens, Stevenson, Stone, Strain, Straniero,
  Stratta, Strauss, Strigin, Sturani, Stuver, Summerscales, Sun, Sutton,
  Swinkels, Szczepa\ifmmode~\acute{n}\else \'{n}\fi{}czyk, Tacca, Talukder,
  Tanner, T\'apai, Tarabrin, Taracchini, Taylor, Theeg, Thirugnanasambandam,
  Thomas, Thomas, Thomas, Thorne, Thorne, Thrane, Tiwari, Tiwari, Tokmakov,
  Tomlinson, Tonelli, Torres, Torrie, T\"oyr\"a, Travasso, Traylor, Trifir\`o,
  Tringali, Trozzo, Tse, Turconi, Tuyenbayev, Ugolini, Unnikrishnan, Urban,
  Usman, Vahlbruch, Vajente, Valdes, Vallisneri, van Bakel, van Beuzekom,
  van~den Brand, Van Den~Broeck, Vander-Hyde, van~der Schaaf, van Heijningen,
  van Veggel, Vardaro, Vass, Vas\'uth, Vaulin, Vecchio, Vedovato, Veitch,
  Veitch, Venkateswara, Verkindt, Vetrano, Vicer\'e, Vinciguerra, Vine, Vinet,
  Vitale, Vo, Vocca, Vorvick, Voss, Vousden, Vyatchanin, Wade, Wade, Wade,
  Waldman, Walker, Wallace, Walsh, Wang, Wang, Wang, Wang, Wang, Ward, Ward,
  Warner, Was, Weaver, Wei, Weinert, Weinstein, Weiss, Welborn, Wen,
  We\ss{}els, Westphal, Wette, Whelan, Whitcomb, White, Whiting, Wiesner,
  Wilkinson, Willems, Williams, Williams, Williamson, Willis, Willke, Wimmer,
  Winkelmann, Winkler, Wipf, Wiseman, Wittel, Woan, Worden, Wright, Wu, Yablon,
  Yakushin, Yam, Yamamoto, Yancey, Yap, Yu, Yvert, Zadro\ifmmode~\dot{z}\else
  \.{z}\fi{}ny, Zangrando, Zanolin, Zendri, Zevin, Zhang, Zhang, Zhang, Zhang,
  Zhao, Zhou, Zhou, Zhu, Zucker, Zuraw, \& Zweizig}]{GW150914}
Abbott, B.~P., Abbott, R., Abbott, T.~D., {et~al.} 2016,
  \href{http://dx.doi.org/10.1103/PhysRevLett.116.061102}{\JournalTitle{Phys.
  Rev. Lett.}, 116, 061102}

\bibitem[{{Abbott} {et~al.}(2017{\natexlab{a}}){Abbott}, {Abbott}, {Abbott},
  {Acernese}, {Ackley}, {Adams}, {Adams}, {Addesso}, {Adhikari}, {Adya}, \&
  et~al.}]{2017Natur.551...85A}
{Abbott}, B.~P., {Abbott}, R., {Abbott}, T.~D., {et~al.} 2017{\natexlab{a}},
  \href{http://dx.doi.org/10.1038/nature24471}{\JournalTitle{\nat}, 551, 85}

\bibitem[{{Abbott} {et~al.}(2017{\natexlab{b}}){Abbott}, {Abbott}, {Abbott},
  {Acernese}, {Ackley}, {Adams}, {Adams}, {Addesso}, {Adhikari}, {Adya}, \&
  et~al.}]{gw170814}
---. 2017{\natexlab{b}},
  \href{http://dx.doi.org/10.1103/PhysRevLett.119.141101}{\JournalTitle{Physical
  Review Letters}, 119, 141101}

\bibitem[{Abbott {et~al.}(2017)Abbott, Abbott, Abbott, Acernese, Ackley, Adams,
  Adams, Addesso, Adhikari, Adya, Affeldt, Afrough, Agarwal, Agathos, Agatsuma,
  Aggarwal, Aguiar, Aiello, Ain, Ajith, Allen, Allen, Allocca, Altin, Amato,
  Ananyeva, Anderson, Anderson, Angelova, Antier, Appert, Arai, Araya, Areeda,
  Arnaud, Arun, Ascenzi, Ashton, Ast, Aston, Astone, Atallah, Aufmuth, Aulbert,
  AultONeal, Austin, Avila-Alvarez, Babak, Bacon, Bader, Bae, Bailes, Baker,
  Baldaccini, Ballardin, Ballmer, Banagiri, Barayoga, Barclay, Barish, Barker,
  Barkett, Barone, Barr, Barsotti, Barsuglia, Barta, Barthelmy, Bartlett,
  Bartos, Bassiri, Basti, Batch, Bawaj, Bayley, Bazzan, B\'ecsy, Beer, Bejger,
  Belahcene, Bell, Berger, Bergmann, Bernuzzi, Bero, Berry, Bersanetti,
  Bertolini, Betzwieser, Bhagwat, Bhandare, Bilenko, Billingsley, Billman,
  Birch, Birney, Birnholtz, Biscans, Biscoveanu, Bisht, Bitossi, Biwer,
  Bizouard, Blackburn, Blackman, Blair, Blair, Blair, Bloemen, Bock, Bode,
  Boer, Bogaert, Bohe, Bondu, Bonilla, Bonnand, Boom, Bork, Boschi, Bose,
  Bossie, Bouffanais, Bozzi, Bradaschia, Brady, Branchesi, Brau, Briant,
  Brillet, Brinkmann, Brisson, Brockill, Broida, Brooks, Brown, Brown, Brunett,
  Buchanan, Buikema, Bulik, Bulten, Buonanno, Buskulic, Buy, Byer, Cabero,
  Cadonati, Cagnoli, Cahillane, Calder\'on~Bustillo, Callister, Calloni, Camp,
  Canepa, Canizares, Cannon, Cao, Cao, Capano, Capocasa, Carbognani, Caride,
  Carney, Carullo, Casanueva~Diaz, Casentini, Caudill, Cavagli\`a, Cavalier,
  Cavalieri, Cella, Cepeda, Cerd\'a-Dur\'an, Cerretani, Cesarini, Chamberlin,
  Chan, Chao, Charlton, Chase, Chassande-Mottin, Chatterjee, Chatziioannou,
  Cheeseboro, Chen, Chen, Chen, Cheng, Chia, Chincarini, Chiummo, Chmiel, Cho,
  Cho, Chow, Christensen, Chu, Chua, Chua, Chung, Chung, Ciani, Ciolfi,
  Cirelli, Cirone, Clara, Clark, Clearwater, Cleva, Cocchieri, Coccia, Cohadon,
  Cohen, Colla, Collette, Cominsky, Constancio, Conti, Cooper, Corban, Corbitt,
  Cordero-Carri\'on, Corley, Cornish, Corsi, Cortese, Costa, Coughlin,
  Coughlin, Coulon, Countryman, Couvares, Covas, Cowan, Coward, Cowart, Coyne,
  Coyne, Creighton, Creighton, Cripe, Crowder, Cullen, Cumming, Cunningham,
  Cuoco, Dal~Canton, D\'alya, Danilishin, D'Antonio, Danzmann, Dasgupta,
  Da~Silva~Costa, Dattilo, Dave, Davier, Davis, Daw, Day, De, DeBra, Degallaix,
  De~Laurentis, Del\'eglise, Del~Pozzo, Demos, Denker, Dent, De~Pietri,
  Dergachev, De~Rosa, DeRosa, De~Rossi, DeSalvo, de~Varona, Devenson,
  Dhurandhar, D\'{\i}az, Dietrich, Di~Fiore, Di~Giovanni, Di~Girolamo,
  Di~Lieto, Di~Pace, Di~Palma, Di~Renzo, Doctor, Dolique, Donovan, Dooley,
  Doravari, Dorrington, Douglas, Dovale~\'Alvarez, Downes, Drago,
  Dreissigacker, Driggers, Du, Ducrot, Dudi, Dupej, Dwyer, Edo, Edwards,
  Effler, Eggenstein, Ehrens, Eichholz, Eikenberry, Eisenstein, Essick,
  Estevez, Etienne, Etzel, Evans, Evans, Factourovich, Fafone, Fair, Fairhurst,
  Fan, Farinon, Farr, Farr, Fauchon-Jones, Favata, Fays, Fee, Fehrmann, Feicht,
  Fejer, Fernandez-Galiana, Ferrante, Ferreira, Ferrini, Fidecaro, Finstad,
  Fiori, Fiorucci, Fishbach, Fisher, Fitz-Axen, Flaminio, Fletcher, Fong, Font,
  Forsyth, Forsyth, Fournier, Frasca, Frasconi, Frei, Freise, Frey, Frey,
  Fries, Fritschel, Frolov, Fulda, Fyffe, Gabbard, Gadre, Gaebel, Gair,
  Gammaitoni, Ganija, Gaonkar, Garcia-Quiros, Garufi, Gateley, Gaudio, Gaur,
  Gayathri, Gehrels, Gemme, Genin, Gennai, George, George, Gergely, Germain,
  Ghonge, Ghosh, Ghosh, Ghosh, Giaime, Giardina, Giazotto, Gill, Glover, Goetz,
  Goetz, Gomes, Goncharov, Gonz\'alez, Gonzalez~Castro, Gopakumar, Gorodetsky,
  Gossan, Gosselin, Gouaty, Grado, Graef, Granata, Grant, Gras, Gray, Greco,
  Green, Gretarsson, Groot, Grote, Grunewald, Gruning, Guidi, Guo, Gupta,
  Gupta, Gushwa, Gustafson, Gustafson, Halim, Hall, Hall, Hamilton, Hammond,
  Haney, Hanke, Hanks, Hanna, Hannam, Hannuksela, Hanson, Hardwick, Harms,
  Harry, Harry, Hart, Haster, Haughian, Healy, Heidmann, Heintze, Heitmann,
  Hello, Hemming, Hendry, Heng, Hennig, Heptonstall, Heurs, Hild, Hinderer, Ho,
  Hoak, Hofman, Holt, Holz, Hopkins, Horst, Hough, Houston, Howell, Hreibi, Hu,
  Huerta, Huet, Hughey, Husa, Huttner, Huynh-Dinh, Indik, Inta, Intini, Isa,
  Isac, Isi, Iyer, Izumi, Jacqmin, Jani, Jaranowski, Jawahar,
  Jim\'enez-Forteza, Johnson, Johnson-McDaniel, Jones, Jones, Jonker, Ju,
  Junker, Kalaghatgi, Kalogera, Kamai, Kandhasamy, Kang, Kanner, Kapadia,
  Karki, Karvinen, Kasprzack, Kastaun, Katolik, Katsavounidis, Katzman, Kaufer,
  Kawabe, K\'ef\'elian, Keitel, Kemball, Kennedy, Kent, Key, Khalili, Khan,
  Khan, Khan, Khazanov, Kijbunchoo, Kim, Kim, Kim, Kim, Kim, Kim, Kimbrell,
  King, King, Kinley-Hanlon, Kirchhoff, Kissel, Kleybolte, Klimenko, Knowles,
  Koch, Koehlenbeck, Koley, Kondrashov, Kontos, Korobko, Korth, Kowalska,
  Kozak, Kr\"amer, Kringel, Krishnan, Kr\'olak, Kuehn, Kumar, Kumar, Kumar,
  Kuo, Kutynia, Kwang, Lackey, Lai, Landry, Lang, Lange, Lantz, Lanza, Larson,
  Lartaux-Vollard, Lasky, Laxen, Lazzarini, Lazzaro, Leaci, Leavey, Lee, Lee,
  Lee, Lee, Lee, Lehmann, Lenon, Leon, Leonardi, Leroy, Letendre, Levin, Li,
  Linker, Littenberg, Liu, Liu, Lo, Lockerbie, London, Lord, Lorenzini,
  Loriette, Lormand, Losurdo, Lough, Lousto, Lovelace, L\"uck, Lumaca,
  Lundgren, Lynch, Ma, Macas, Macfoy, Machenschalk, MacInnis, Macleod, Maga\~na
  Hernandez, Maga\~na Sandoval, Maga\~na Zertuche, Magee, Majorana, Maksimovic,
  Man, Mandic, Mangano, Mansell, Manske, Mantovani, Marchesoni, Marion,
  M\'arka, M\'arka, Markakis, Markosyan, Markowitz, Maros, Marquina, Marsh,
  Martelli, Martellini, Martin, Martin, Martynov, Marx, Mason, Massera,
  Masserot, Massinger, Masso-Reid, Mastrogiovanni, Matas, Matichard, Matone,
  Mavalvala, Mazumder, McCarthy, McClelland, McCormick, McCuller, McGuire,
  McIntyre, McIver, McManus, McNeill, McRae, McWilliams, Meacher, Meadors,
  Mehmet, Meidam, Mejuto-Villa, Melatos, Mendell, Mercer, Merilh, Merzougui,
  Meshkov, Messenger, Messick, Metzdorff, Meyers, Miao, Michel, Middleton,
  Mikhailov, Milano, Miller, Miller, Miller, Millhouse, Milovich-Goff,
  Minazzoli, Minenkov, Ming, Mishra, Mitra, Mitrofanov, Mitselmakher,
  Mittleman, Moffa, Moggi, Mogushi, Mohan, Mohapatra, Molina, Montani, Moore,
  Moraru, Moreno, Morisaki, Morriss, Mours, Mow-Lowry, Mueller, Muir,
  Mukherjee, Mukherjee, Mukherjee, Mukund, Mullavey, Munch, Mu\~niz, Muratore,
  Murray, Nagar, Napier, Nardecchia, Naticchioni, Nayak, Neilson, Nelemans,
  Nelson, Nery, Neunzert, Nevin, Newport, Newton, Ng, Nguyen, Nguyen, Nichols,
  Nielsen, Nissanke, Nitz, Noack, Nocera, Nolting, North, Nuttall, Oberling,
  O'Dea, Ogin, Oh, Oh, Ohme, Okada, Oliver, Oppermann, Oram, O'Reilly,
  Ormiston, Ortega, O'Shaughnessy, Ossokine, Ottaway, Overmier, Owen, Pace,
  Page, Page, Pai, Pai, Palamos, Palashov, Palomba, Pal-Singh, Pan, Pan, Pang,
  Pang, Pankow, Pannarale, Pant, Paoletti, Paoli, Papa, Parida, Parker,
  Pascucci, Pasqualetti, Passaquieti, Passuello, Patil, Patricelli, Pearlstone,
  Pedraza, Pedurand, Pekowsky, Pele, Penn, Perez, Perreca, Perri, Pfeiffer,
  Phelps, Piccinni, Pichot, Piergiovanni, Pierro, Pillant, Pinard, Pinto,
  Pirello, Pitkin, Poe, Poggiani, Popolizio, Porter, Post, Powell, Prasad,
  Pratt, Pratten, Predoi, Prestegard, Prijatelj, Principe, Privitera, Prix,
  Prodi, Prokhorov, Puncken, Punturo, Puppo, P\"urrer, Qi, Quetschke, Quintero,
  Quitzow-James, Raab, Rabeling, Radkins, Raffai, Raja, Rajan, Rajbhandari,
  Rakhmanov, Ramirez, Ramos-Buades, Rapagnani, Raymond, Razzano, Read,
  Regimbau, Rei, Reid, Reitze, Ren, Reyes, Ricci, Ricker, Rieger, Riles, Rizzo,
  Robertson, Robie, Robinet, Rocchi, Rolland, Rollins, Roma, Romano, Romano,
  Romel, Romie, Rosi\ifmmode~\acute{n}\else \'{n}\fi{}ska, Ross, Rowan,
  R\"udiger, Ruggi, Rutins, Ryan, Sachdev, Sadecki, Sadeghian, Sakellariadou,
  Salconi, Saleem, Salemi, Samajdar, Sammut, Sampson, Sanchez, Sanchez,
  Sanchis-Gual, Sandberg, Sanders, Sassolas, Sathyaprakash, Saulson, Sauter,
  Savage, Sawadsky, Schale, Scheel, Scheuer, Schmidt, Schmidt, Schnabel,
  Schofield, Sch\"onbeck, Schreiber, Schuette, Schulte, Schutz, Schwalbe,
  Scott, Scott, Seidel, Sellers, Sengupta, Sentenac, Sequino, Sergeev,
  Shaddock, Shaffer, Shah, Shahriar, Shaner, Shao, Shapiro, Shawhan, Sheperd,
  Shoemaker, Shoemaker, Siellez, Siemens, Sieniawska, Sigg, Silva, Singer,
  Singh, Singhal, Sintes, Slagmolen, Smith, Smith, Smith, Somala, Son,
  Sonnenberg, Sorazu, Sorrentino, Souradeep, Spencer, Srivastava, Staats,
  Staley, Steinke, Steinlechner, Steinlechner, Steinmeyer, Stevenson, Stone,
  Stops, Strain, Stratta, Strigin, Strunk, Sturani, Stuver, Summerscales, Sun,
  Sunil, Suresh, Sutton, Swinkels, Szczepa\ifmmode~\acute{n}\else
  \'{n}\fi{}czyk, Tacca, Tait, Talbot, Talukder, Tanner, T\'apai, Taracchini,
  Tasson, Taylor, Taylor, Tewari, Theeg, Thies, Thomas, Thomas, Thomas, Thorne,
  Thorne, Thrane, Tiwari, Tiwari, Tokmakov, Toland, Tonelli, Tornasi,
  Torres-Forn\'e, Torrie, T\"oyr\"a, Travasso, Traylor, Trinastic, Tringali,
  Trozzo, Tsang, Tse, Tso, Tsukada, Tsuna, Tuyenbayev, Ueno, Ugolini,
  Unnikrishnan, Urban, Usman, Vahlbruch, Vajente, Valdes, Vallisneri, van
  Bakel, van Beuzekom, van~den Brand, Van Den~Broeck, Vander-Hyde, van~der
  Schaaf, van Heijningen, van Veggel, Vardaro, Varma, Vass, Vas\'uth, Vecchio,
  Vedovato, Veitch, Veitch, Venkateswara, Venugopalan, Verkindt, Vetrano,
  Vicer\'e, Viets, Vinciguerra, Vine, Vinet, Vitale, Vo, Vocca, Vorvick,
  Vyatchanin, Wade, Wade, Wade, Walet, Walker, Wallace, Walsh, Wang, Wang,
  Wang, Wang, Wang, Ward, Warner, Was, Watchi, Weaver, Wei, Weinert, Weinstein,
  Weiss, Wen, Wessel, We\ss{}els, Westerweck, Westphal, Wette, Whelan,
  Whitcomb, Whiting, Whittle, Wilken, Williams, Williams, Williamson, Willis,
  Willke, Wimmer, Winkler, Wipf, Wittel, Woan, Woehler, Wofford, Wong, Worden,
  Wright, Wu, Wysocki, Xiao, Yamamoto, Yancey, Yang, Yap, Yazback, Yu, Yu,
  Yvert, Zadro\ifmmode~\dot{z}\else \.{z}\fi{}ny, Zanolin, Zelenova, Zendri,
  Zevin, Zhang, Zhang, Zhang, Zhang, Zhao, Zhou, Zhou, Zhu, Zhu, Zimmerman,
  Zucker, \& Zweizig}]{ligobns}
Abbott, B.~P., Abbott, R., Abbott, T.~D., {et~al.} 2017,
  \href{http://dx.doi.org/10.1103/PhysRevLett.119.161101}{\JournalTitle{Phys.
  Rev. Lett.}, 119, 161101}

\bibitem[{{Abbott} {et~al.}(2016){Abbott}, {Abdalla}, {Allam}, {Amara},
  {Annis}, {Armstrong}, {Bacon}, {Banerji}, {Bauer}, {Baxter}, {Becker},
  {Benoit-L{\'e}vy}, {Bernstein}, {Bernstein}, {Bertin}, {Blazek}, {Bonnett},
  {Bridle}, {Brooks}, {Bruderer}, {Buckley-Geer}, {Burke}, {Busha}, {Capozzi},
  {Carnero Rosell}, {Carrasco Kind}, {Carretero}, {Castander}, {Chang},
  {Clampitt}, {Crocce}, {Cunha}, {D'Andrea}, {da Costa}, {Das}, {DePoy},
  {Desai}, {Diehl}, {Dietrich}, {Dodelson}, {Doel}, {Drlica-Wagner},
  {Efstathiou}, {Eifler}, {Erickson}, {Estrada}, {Evrard}, {Fausti Neto},
  {Fernandez}, {Finley}, {Flaugher}, {Fosalba}, {Friedrich}, {Frieman},
  {Gangkofner}, {Garcia-Bellido}, {Gaztanaga}, {Gerdes}, {Gruen}, {Gruendl},
  {Gutierrez}, {Hartley}, {Hirsch}, {Honscheid}, {Huff}, {Jain}, {James},
  {Jarvis}, {Kacprzak}, {Kent}, {Kirk}, {Krause}, {Kravtsov}, {Kuehn},
  {Kuropatkin}, {Kwan}, {Lahav}, {Leistedt}, {Li}, {Lima}, {Lin}, {MacCrann},
  {March}, {Marshall}, {Martini}, {McMahon}, {Melchior}, {Miller}, {Miquel},
  {Mohr}, {Neilsen}, {Nichol}, {Nicola}, {Nord}, {Ogando}, {Palmese}, {Peiris},
  {Plazas}, {Refregier}, {Roe}, {Romer}, {Roodman}, {Rowe}, {Rykoff}, {Sabiu},
  {Sadeh}, {Sako}, {Samuroff}, {Sanchez}, {S{\'a}nchez}, {Seo},
  {Sevilla-Noarbe}, {Sheldon}, {Smith}, {Soares-Santos}, {Sobreira}, {Suchyta},
  {Swanson}, {Tarle}, {Thaler}, {Thomas}, {Troxel}, {Vikram}, {Walker},
  {Wechsler}, {Weller}, {Zhang}, {Zuntz}, \& {Dark Energy Survey
  Collaboration}}]{2016PhRvD..94b2001A}
{Abbott}, T., {Abdalla}, F.~B., {Allam}, S., {et~al.} 2016,
  \href{http://dx.doi.org/10.1103/PhysRevD.94.022001}{\JournalTitle{\prd}, 94,
  022001}

\bibitem[{{Abbott} {et~al.}(2018){Abbott}, {Abdalla}, {Allam}, {Amara},
  {Annis}, {Asorey}, {Avila}, {Ballester}, {Banerji}, {Barkhouse}, {Baruah},
  {Baumer}, {Bechtol}, {Becker}, {Benoit-L{\'e}vy}, {Bernstein}, {Bertin},
  {Blazek}, {Bocquet}, {Brooks}, {Brout}, {Buckley-Geer}, {Burke}, {Busti},
  {Campisano}, {Cardiel-Sas}, {arnero Rosell}, {Carrasco Kind}, {Carretero},
  {Castander}, {Cawthon}, {Chang}, {Conselice}, {Costa}, {Crocce}, {Cunha},
  {D'Andrea}, {da Costa}, {Das}, {Daues}, {Davis}, {Davis}, {De Vicente},
  {DePoy}, {DeRose}, {Desai}, {Diehl}, {Dietrich}, {Dodelson}, {Doel},
  {Drlica-Wagner}, {Eifler}, {Elliott}, {Evrard}, {Farahi}, {Fausti Neto},
  {Fernandez}, {Finley}, {Fitzpatrick}, {Flaugher}, {Foley}, {Fosalba},
  {Friedel}, {Frieman}, {Garc{\'{\i}}a-Bellido}, {Gaz tanaga}, {Gerdes},
  {Giannantonio}, {Gill}, {Glazebrook}, {Goldstein}, {Gower}, {Gruen},
  {Gruendl}, {Gschwend}, {Gupta}, {Gutierrez}, {Hamilton}, {Hartley}, {Hinton},
  {Hislop}, {Hollowood}, {Honscheid}, {Hoyle}, {Huterer}, {Jain}, {James},
  {Jeltema}, {Johnson}, {Johnson}, {Juneau}, {Kacpr zak}, {Kent}, {Khullar},
  {Klein}, {Kovacs}, {Koziol}, {Krause}, {Kremin}, {Kron}, {Kuehn}, {Kuhlmann},
  {Kuropatkin}, {Lahav}, {Lasker}, {Li}, {Li}, {Liddle}, {Lima}, {Lin},
  {L{\'o}pez-Reyes}, {MacCrann}, {Maia}, {Maloney}, {Manera}, {March},
  {Marriner}, {Marshall}, {Martini}, {McClintock}, {McKay}, {McMahon},
  {Melchior}, {Menanteau}, {Miller}, {Miquel}, {Mohr}, {Morganson}, {Mould},
  {Neilsen}, {Nichol}, {Nidever}, {Nikutta}, {Nogueira}, {Nord}, {Nugent},
  {Nunes}, {Ogando}, {Old}, {Olsen}, {Pace}, {Palmese}, {Paz-Chinch{\'o}n},
  {Peiris}, {Percival}, {Petravick}, {Plazas}, {Poh}, {Pond}, {Por redon},
  {Pujol}, {Refregier}, {Reil}, {Ricker}, {Rollins}, {Romer}, {Roodman},
  {Rooney}, {Ross}, {Rykoff}, {Sako}, {Sanchez}, {Sanchez}, {Santiago}, {Saro},
  {Scarpine}, {Scolnic}, {Scott}, {Serrano}, {Sevilla-Noarbe}, {Sheldon},
  {Shipp}, {Silveira}, {Smith}, {Smith}, {Smith}, {Soares-Santos}, {Sobre ira},
  {Song}, {Stebbins}, {Suchyta}, {Sullivan}, {Swanson}, {Tarle}, {Thaler},
  {Thomas}, {Thomas}, {Troxel}, {Tucker}, {Vikram}, {Vivas}, {Wal ker},
  {Wechsler}, {Weller}, {Wester}, {Wolf}, {Wu}, {Yanny}, {Zenteno}, {Zhang}, \&
  {Zuntz}}]{dr1}
{Abbott}, T.~M.~C., {Abdalla}, F.~B., {Allam}, S., {et~al.} 2018,
  \JournalTitle{ArXiv e-prints},
  \href{http://arxiv.org/abs/1801.03181}{{\sffamily arXiv:1801.03181
  [astro-ph.IM]}}

\bibitem[{{Allen}(1976)}]{allen}
{Allen}, D.~A. 1976,
  \href{http://dx.doi.org/10.1093/mnras/174.1.29P}{\JournalTitle{\mnras}, 174,
  29P}

\bibitem[{{Arcavi} {et~al.}(2017){Arcavi}, {Hosseinzadeh}, {Howell}, {McCully},
  {Poznanski}, {Kasen}, {Barnes}, {Zaltzman}, {Vasylyev}, {Maoz}, \&
  {Valenti}}]{arcavi}
{Arcavi}, I., {Hosseinzadeh}, G., {Howell}, D.~A., {et~al.} 2017,
  \href{http://dx.doi.org/10.1038/nature24291}{\JournalTitle{\nat}, 551, 64}

\bibitem[{{Arnouts} {et~al.}(1999){Arnouts}, {Cristiani}, {Moscardini},
  {Matarrese}, {Lucchin}, {Fontana}, \& {Giallongo}}]{arnouts}
{Arnouts}, S., {Cristiani}, S., {Moscardini}, L., {et~al.} 1999,
  \href{http://dx.doi.org/10.1046/j.1365-8711.1999.02978.x}{\JournalTitle{\mnras},
  310, 540}

\bibitem[{{Bayliss} {et~al.}(2016){Bayliss}, {Ruel}, {Stubbs}, {Allen},
  {Applegate}, {Ashby}, {Bautz}, {Benson}, {Bleem}, {Bocquet}, {Brodwin},
  {Capasso}, {Carlstrom}, {Chang}, {Chiu}, {Cho}, {Clocchiatti}, {Crawford},
  {Crites}, {de Haan}, {Desai}, {Dietrich}, {Dobbs}, {Doucouliagos}, {Foley},
  {Forman}, {Garmire}, {George}, {Gladders}, {Gonzalez}, {Gupta}, {Halverson},
  {Hlavacek-Larrondo}, {Hoekstra}, {Holder}, {Holzapfel}, {Hou}, {Hrubes},
  {Huang}, {Jones}, {Keisler}, {Knox}, {Lee}, {Leitch}, {von der Linden},
  {Luong-Van}, {Mantz}, {Marrone}, {McDonald}, {McMahon}, {Meyer}, {Mocanu},
  {Mohr}, {Murray}, {Padin}, {Pryke}, {Rapetti}, {Reichardt}, {Rest}, {Ruhl},
  {Saliwanchik}, {Saro}, {Sayre}, {Schaffer}, {Schrabback}, {Shirokoff},
  {Song}, {Spieler}, {Stalder}, {Stanford}, {Staniszewski}, {Stark}, {Story},
  {Vanderlinde}, {Vieira}, {Vikhlinin}, {Williamson}, \&
  {Zenteno}}]{2016ApJS..227....3B}
{Bayliss}, M.~B., {Ruel}, J., {Stubbs}, C.~W., {et~al.} 2016,
  \href{http://dx.doi.org/10.3847/0067-0049/227/1/3}{\JournalTitle{\apjs}, 227,
  3}

\bibitem[{{Ben{\'{\i}}tez}(2000)}]{benitez}
{Ben{\'{\i}}tez}, N. 2000,
  \href{http://dx.doi.org/10.1086/308947}{\JournalTitle{\apj}, 536, 571}

\bibitem[{{Bertin} \& {Arnouts}(1996)}]{sextractor}
{Bertin}, E., \& {Arnouts}, S. 1996,
  \href{http://dx.doi.org/10.1051/aas:1996164}{\JournalTitle{Astronomy and
  Astrophysics Supplement}, 117, 393}

\bibitem[{{Bilicki} {et~al.}(2018){Bilicki}, {Hoekstra}, {Brown}, {Amaro},
  {Blake}, {Cavuoti}, {de Jong}, {Georgiou}, {Hildebrandt}, {Wolf}, {Amon},
  {Brescia}, {Brough}, {Costa-Duarte}, {Erben}, {Glazebrook}, {Grado},
  {Heymans}, {Jarrett}, {Joudaki}, {Kuijken}, {Longo}, {Napolitano},
  {Parkinson}, {Vellucci}, {Kleijn}, \& {Wang}}]{bilicki}
{Bilicki}, M., {Hoekstra}, H., {Brown}, M.~J.~I., {et~al.} 2018,
  \href{http://dx.doi.org/10.1051/0004-6361/201731942}{\JournalTitle{\aap},
  616, A69}

\bibitem[{Bird {et~al.}(2016)Bird, Cholis, Mu\~noz, Ali-Ha\"{\i}moud,
  Kamionkowski, Kovetz, Raccanelli, \& Riess}]{PhysRevLett.116.201301}
Bird, S., Cholis, I., Mu\~noz, J.~B., {et~al.} 2016,
  \href{http://dx.doi.org/10.1103/PhysRevLett.116.201301}{\JournalTitle{Phys.
  Rev. Lett.}, 116, 201301}

\bibitem[{{Bonnett} {et~al.}(2016){Bonnett}, {Troxel}, {Hartley}, {Amara},
  {Leistedt}, {Becker}, {Bernstein}, {Bridle}, {Bruderer}, {Busha}, {Carrasco
  Kind}, {Childress}, {Castander}, {Chang}, {Crocce}, {Davis}, {Eifler},
  {Frieman}, {Gangkofner}, {Gaztanaga}, {Glazebrook}, {Gruen}, {Kacprzak},
  {King}, {Kwan}, {Lahav}, {Lewis}, {Lidman}, {Lin}, {MacCrann}, {Miquel},
  {O'Neill}, {Palmese}, {Peiris}, {Refregier}, {Rozo}, {Rykoff}, {Sadeh},
  {S{\'a}nchez}, {Sheldon}, {Uddin}, {Wechsler}, {Zuntz}, {Abbott}, {Abdalla},
  {Allam}, {Armstrong}, {Banerji}, {Bauer}, {Benoit-L{\'e}vy}, {Bertin},
  {Brooks}, {Buckley-Geer}, {Burke}, {Capozzi}, {Carnero Rosell}, {Carretero},
  {Cunha}, {D'Andrea}, {da Costa}, {DePoy}, {Desai}, {Diehl}, {Dietrich},
  {Doel}, {Fausti Neto}, {Fernandez}, {Flaugher}, {Fosalba}, {Gerdes},
  {Gruendl}, {Honscheid}, {Jain}, {James}, {Jarvis}, {Kim}, {Kuehn},
  {Kuropatkin}, {Li}, {Lima}, {Maia}, {March}, {Marshall}, {Martini},
  {Melchior}, {Miller}, {Neilsen}, {Nichol}, {Nord}, {Ogando}, {Plazas},
  {Reil}, {Romer}, {Roodman}, {Sako}, {Sanchez}, {Santiago}, {Smith},
  {Soares-Santos}, {Sobreira}, {Suchyta}, {Swanson}, {Tarle}, {Thaler},
  {Thomas}, {Vikram}, {Walker}, \& {Dark Energy Survey
  Collaboration}}]{bonnett}
{Bonnett}, C., {Troxel}, M.~A., {Hartley}, W., {et~al.}
  \href{http://dx.doi.org/10.1103/PhysRevD.94.042005}{2016, 94, 042005}

\bibitem[{{Bruzual} \& {Charlot}(2003)}]{bc03}
{Bruzual}, G., \& {Charlot}, S. 2003,
  \href{http://dx.doi.org/10.1046/j.1365-8711.2003.06897.x}{\JournalTitle{\mnras},
  344, 1000}

\bibitem[{{Cao} {et~al.}(2018){Cao}, {Lu}, \& {Zhao}}]{2018MNRAS.474.4997C}
{Cao}, L., {Lu}, Y., \& {Zhao}, Y. 2018,
  \href{http://dx.doi.org/10.1093/mnras/stx3087}{\JournalTitle{\mnras}, 474,
  4997}

\bibitem[{{Chabrier}(2003)}]{chabrier}
{Chabrier}, G. 2003,
  \href{http://dx.doi.org/10.1086/376392}{\JournalTitle{\pasp}, 115, 763}

\bibitem[{{Chambers} {et~al.}(2016){Chambers}, {Magnier}, {Metcalfe},
  {Flewelling}, {Huber}, {Waters}, {Denneau}, {Draper}, {Farrow}, {Finkbeiner},
  {Holmberg}, {Koppenhoefer}, {Price}, {Saglia}, {Schlafly}, {Smartt},
  {Sweeney}, {Wainscoat}, {Burgett}, {Grav}, {Heasley}, {Hodapp}, {Jedicke},
  {Kaiser}, {Kudritzki}, {Luppino}, {Lupton}, {Monet}, {Morgan}, {Onaka},
  {Stubbs}, {Tonry}, {Banados}, {Bell}, {Bender}, {Bernard}, {Botticella},
  {Casertano}, {Chastel}, {Chen}, {Chen}, {Cole}, {Deacon}, {Frenk},
  {Fitzsimmons}, {Gezari}, {Goessl}, {Goggia}, {Goldman}, {Grebel}, {Hambly},
  {Hasinger}, {Heavens}, {Heckman}, {Henderson}, {Henning}, {Holman}, {Hopp},
  {Ip}, {Isani}, {Keyes}, {Koekemoer}, {Kotak}, {Long}, {Lucey}, {Liu},
  {Martin}, {McLean}, {Morganson}, {Murphy}, {Nieto- Santisteban}, {Norberg},
  {Peacock}, {Pier}, {Postman}, {Primak}, {Rae}, {Rest}, {Riess}, {Riffeser},
  {Rix}, {Roser}, {Schilbach}, {Schultz}, {Scolnic}, {Szalay}, {Seitz},
  {Shiao}, {Small}, {Smith}, {Soderblom}, {Taylor}, {Thakar}, {Thiel},
  {Thilker}, {Urata}, {Valenti}, {Walter}, {Watters}, {Werner}, {White}, {Wood-
  Vasey}, \& {Wyse}}]{panstarrs}
{Chambers}, K.~C., {Magnier}, E.~A., {Metcalfe}, N., {et~al.} 2016,
  \JournalTitle{arXiv e-prints}, arXiv:1612.05560

\bibitem[{{Chen} {et~al.}(2018){Chen}, {Fishbach}, \& {Holz}}]{chen17}
{Chen}, H.-Y., {Fishbach}, M., \& {Holz}, D.~E. 2018,
  \href{http://dx.doi.org/10.1038/s41586-018-0606-0}{\JournalTitle{\nat}, 562,
  545}

\bibitem[{{Chen} \& {Holz}(2016)}]{chen12}
{Chen}, H.-Y., \& {Holz}, D.~E. 2016, \JournalTitle{arXiv e-prints},
  arXiv:1612.01471

\bibitem[{{Clesse} \& {Garc{\'\i}a-Bellido}(2017)}]{2017PDU....15..142C}
{Clesse}, S., \& {Garc{\'\i}a-Bellido}, J. 2017,
  \href{http://dx.doi.org/10.1016/j.dark.2016.10.002}{\JournalTitle{Physics of
  the Dark Universe}, 15, 142}

\bibitem[{{Clesse} \& {Garc{\'{\i}}a-Bellido}(2018)}]{pbh2}
{Clesse}, S., \& {Garc{\'{\i}}a-Bellido}, J. 2018,
  \href{http://dx.doi.org/10.1016/j.dark.2018.08.004}{\JournalTitle{Physics of
  the Dark Universe}, 22, 137}

\bibitem[{{Colless} {et~al.}(2001){Colless}, {Dalton}, {Maddox}, {Sutherland},
  {Norberg}, {Cole}, {Bland-Hawthorn}, {Bridges}, {Cannon}, {Collins}, {Couch},
  {Cross}, {Deeley}, {De Propris}, {Driver}, {Efstathiou}, {Ellis}, {Frenk},
  {Glazebrook}, {Jackson}, {Lahav}, {Lewis}, {Lumsden}, {Madgwick}, {Peacock},
  {Peterson}, {Price}, {Seaborne}, \& {Taylor}}]{2001MNRAS.328.1039C}
{Colless}, M., {Dalton}, G., {Maddox}, S., {et~al.} 2001,
  \href{http://dx.doi.org/10.1046/j.1365-8711.2001.04902.x}{\JournalTitle{\mnras},
  328, 1039}

\bibitem[{Coulter {et~al.}(2017)Coulter, Foley, Kilpatrick, Drout, Piro,
  Shappee, Siebert, Simon, Ulloa, Kasen, Madore, Murguia-Berthier, Pan,
  Prochaska, Ramirez-Ruiz, Rest, \& Rojas-Bravo}]{Coulter1556}
Coulter, D.~A., Foley, R.~J., Kilpatrick, C.~D., {et~al.} 2017,
  \href{http://dx.doi.org/10.1126/science.aap9811}{\JournalTitle{Science}, 358,
  1556}

\bibitem[{{Cowperthwaite} {et~al.}(2016){Cowperthwaite}, {Berger},
  {Soares-Santos}, {Annis}, {Brout}, {Brown}, {Buckley-Geer}, {Cenko}, {Chen},
  {Chornock}, {Diehl}, {Doctor}, {Drlica-Wagner}, {Drout}, {Farr}, {Finley},
  {Foley}, {Fong}, {Fox}, {Frieman}, {Garcia-Bellido}, {Gill}, {Gruendl},
  {Herner}, {Holz}, {Kasen}, {Kessler}, {Lin}, {Margutti}, {Marriner},
  {Matheson}, {Metzger}, {Neilsen}, {Quataert}, {Rest}, {Sako}, {Scolnic},
  {Smith}, {Sobreira}, {Strampelli}, {Villar}, {Walker}, {Wester}, {Williams},
  {Yanny}, {Abbott}, {Abdalla}, {Allam}, {Armstrong}, {Bechtol},
  {Benoit-L{\'e}vy}, {Bertin}, {Brooks}, {Burke}, {Carnero Rosell}, {Carrasco
  Kind}, {Carretero}, {Castander}, {Cunha}, {D'Andrea}, {da Costa}, {Desai},
  {Dietrich}, {Evrard}, {Fausti Neto}, {Fosalba}, {Gerdes}, {Giannantonio},
  {Goldstein}, {Gruen}, {Gutierrez}, {Honscheid}, {James}, {Johnson},
  {Johnson}, {Krause}, {Kuehn}, {Kuropatkin}, {Lima}, {Maia}, {Marshall},
  {Menanteau}, {Miquel}, {Mohr}, {Nichol}, {Nord}, {Ogando}, {Plazas}, {Reil},
  {Romer}, {Sanchez}, {Scarpine}, {Sevilla-Noarbe}, {Smith}, {Suchyta},
  {Tarle}, {Thomas}, {Thomas}, {Tucker}, {Weller}, \& {DES
  Collaboration}}]{2016ApJ...826L..29C}
{Cowperthwaite}, P.~S., {Berger}, E., {Soares-Santos}, M., {et~al.} 2016,
  \href{http://dx.doi.org/10.3847/2041-8205/826/2/L29}{\JournalTitle{\apjl},
  826, L29}

\bibitem[{{D{\'a}lya} {et~al.}(2018){D{\'a}lya}, {Galg{\'o}czi}, {Dobos},
  {Frei}, {Heng}, {Macas}, {Messenger}, {Raffai}, \& {de Souza}}]{glade}
{D{\'a}lya}, G., {Galg{\'o}czi}, G., {Dobos}, L., {et~al.} 2018,
  \href{http://dx.doi.org/10.1093/mnras/sty1703}{\JournalTitle{\mnras}, 479,
  2374}

\bibitem[{{Dark Energy Survey Collaboration} {et~al.}(2016){Dark Energy Survey
  Collaboration}, {Abbott}, {Abdalla}, {Aleksi{\'c}}, {Allam}, {Amara},
  {Bacon}, {Balbinot}, {Banerji}, {Bechtol}, {Benoit-L{\'e}vy}, {Bernstein},
  {Bertin}, {Blazek}, {Bonnett}, {Bridle}, {Brooks}, {Brunner}, {Buckley-Geer},
  {Burke}, {Caminha}, {Capozzi}, {Carlsen}, {Carnero-Rosell}, {Carollo},
  {Carrasco-Kind}, {Carretero}, {Castander}, {Clerkin}, {Collett}, {Conselice},
  {Crocce}, {Cunha}, {D'Andrea}, {da Costa}, {Davis}, {Desai}, {Diehl},
  {Dietrich}, {Dodelson}, {Doel}, {Drlica-Wagner}, {Estrada}, {Etherington},
  {Evrard}, {Fabbri}, {Finley}, {Flaugher}, {Foley}, {Fosalba}, {Frieman},
  {Garc{\'{\i}}a-Bellido}, {Gaztanaga}, {Gerdes}, {Giannantonio}, {Goldstein},
  {Gruen}, {Gruendl}, {Guarnieri}, {Gutierrez}, {Hartley}, {Honscheid}, {Jain},
  {James}, {Jeltema}, {Jouvel}, {Kessler}, {King}, {Kirk}, {Kron}, {Kuehn},
  {Kuropatkin}, {Lahav}, {Li}, {Lima}, {Lin}, {Maia}, {Makler}, {Manera},
  {Maraston}, {Marshall}, {Martini}, {McMahon}, {Melchior}, {Merson}, {Miller},
  {Miquel}, {Mohr}, {Morice-Atkinson}, {Naidoo}, {Neilsen}, {Nichol}, {Nord},
  {Ogando}, {Ostrovski}, {Palmese}, {Papadopoulos}, {Peiris}, {Peoples},
  {Percival}, {Plazas}, {Reed}, {Refregier}, {Romer}, {Roodman}, {Ross},
  {Rozo}, {Rykoff}, {Sadeh}, {Sako}, {S{\'a}nchez}, {Sanchez}, {Santiago},
  {Scarpine}, {Schubnell}, {Sevilla-Noarbe}, {Sheldon}, {Smith}, {Smith},
  {Soares-Santos}, {Sobreira}, {Soumagnac}, {Suchyta}, {Sullivan}, {Swanson},
  {Tarle}, {Thaler}, {Thomas}, {Thomas}, {Tucker}, {Vieira}, {Vikram},
  {Walker}, {Wechsler}, {Weller}, {Wester}, {Whiteway}, {Wilcox}, {Yanny},
  {Zhang}, \& {Zuntz}}]{2016MNRAS.460.1270D}
{Dark Energy Survey Collaboration}, {Abbott}, T., {Abdalla}, F.~B., {et~al.}
  2016, \href{http://dx.doi.org/10.1093/mnras/stw641}{\JournalTitle{\mnras},
  460, 1270}

\bibitem[{{De Vicente} {et~al.}(2016){De Vicente}, {S{\'a}nchez}, \&
  {Sevilla-Noarbe}}]{dnf}
{De Vicente}, J., {S{\'a}nchez}, E., \& {Sevilla-Noarbe}, I. 2016,
  \href{http://dx.doi.org/10.1093/mnras/stw857}{\JournalTitle{\mnras}, 459,
  3078}

\bibitem[{{Del Pozzo}(2012)}]{delpozzo}
{Del Pozzo}, W. 2012,
  \href{http://dx.doi.org/10.1103/PhysRevD.86.043011}{\JournalTitle{\prd}, 86,
  043011}

\bibitem[{{Doctor} {et~al.}(2018){Doctor}, {Kessler}, {Herner}, {Palmese},
  {Soares-Santos}, {Annis}, {Brout}, {Holz}, {Sako}, {Rest}, {Cowperthwaite},
  {Berger}, {Foley}, {Conselice}, {Gill}, {Allam}, {Balbinot}, {Butler},
  {Chen}, {Chornock}, {Cook}, {Diehl}, {Farr}, {Fong}, {Frieman}, {Fryer},
  {Garc\textbackslash{\'\i}a-Bellido}, {Margutti}, {Marshall}, {Matheson},
  {Metzger}, {Nicholl}, {Paz-Chinch\textbackslash{\'o}n}, {Salim}, {Sauseda},
  {Secco}, {Smith}, {Smith}, {Vivas}, {Tucker}, {Abbott}, {Avila}, {Bechtol},
  {Bertin}, {Brooks}, {Buckley-Geer}, {Burke}, {Carnero Rosell}, {Carrasco
  Kind}, {Carretero}, {Castander}, {D'Andrea}, {da Costa}, {De Vicente},
  {Desai}, {Doel}, {Flaugher}, {Fosalba}, {Gaztanaga}, {Gerdes}, {Goldstein},
  {Gruen}, {Gruendl}, {Gutierrez}, {Hartley}, {Hollowood}, {Honscheid},
  {Hoyle}, {James}, {Jeltema}, {Kent}, {Kuehn}, {Kuropatkin}, {Lahav}, {Lima},
  {Maia}, {March}, {Menanteau}, {Miller}, {Miquel}, {Neilsen}, {Nord},
  {Ogando}, {Plazas}, {Roodman}, {Sanchez}, {Scarpine}, {Schindler},
  {Schubnell}, {Serrano}, {Sevilla-Noarbe}, {Smith}, {Sobreira}, {Suchyta},
  {Swanson}, {Tarle}, {Thomas}, {Walker}, \& {Wester}}]{doctor}
{Doctor}, Z., {Kessler}, R., {Herner}, K., {et~al.} 2018, \JournalTitle{arXiv
  e-prints}, arXiv:1812.01579

\bibitem[{{Drlica-Wagner} {et~al.}(2017){Drlica-Wagner}, {Sevilla-Noarbe},
  {Rykoff}, \& {et al.}}]{firstyear}
{Drlica-Wagner}, A., {Sevilla-Noarbe}, I., {Rykoff}, E.~S., \& {et al.} 2017,
  \JournalTitle{submitted to PRD}

\bibitem[{{Erben} {et~al.}(2013){Erben}, {Hildebrandt}, {Miller}, {van
  Waerbeke}, {Heymans}, {Hoekstra}, {Kitching}, {Mellier}, {Benjamin}, {Blake},
  {Bonnett}, {Cordes}, {Coupon}, {Fu}, {Gavazzi}, {Gillis}, {Grocutt}, {Gwyn},
  {Holhjem}, {Hudson}, {Kilbinger}, {Kuijken}, {Milkeraitis}, {Rowe},
  {Schrabback}, {Semboloni}, {Simon}, {Smit}, {Toader}, {Vafaei}, {van Uitert},
  \& {Velander}}]{Erben13}
{Erben}, T., {Hildebrandt}, H., {Miller}, L., {et~al.} 2013,
  \href{http://dx.doi.org/10.1093/mnras/stt928}{\JournalTitle{\mnras}, 433,
  2545}

\bibitem[{{Etherington} {et~al.}(2017){Etherington}, {Thomas}, {Maraston},
  {Sevilla-Noarbe}, {Bechtol}, {Pforr}, {Pellegrini}, {Gschwend}, {Carnero
  Rosell}, {Maia}, {da Costa}, {Benoit-L{\'e}vy}, {Swanson}, {Hartley},
  {Abbott}, {Abdalla}, {Allam}, {Bernstein}, {Bertin}, {Brooks},
  {Buckley-Geer}, {Carrasco Kind}, {Carretero}, {Castander}, {Crocce}, {Cunha},
  {Desai}, {Doel}, {Eifler}, {Evrard}, {Fausti Neto}, {Finley}, {Flaugher},
  {Fosalba}, {Frieman}, {Gerdes}, {Gruen}, {Gruendl}, {Gutierrez}, {Honscheid},
  {James}, {Kuehn}, {Kuropatkin}, {Lahav}, {Lima}, {Martini}, {Melchior},
  {Miquel}, {Mohr}, {Nord}, {Ogando}, {Plazas}, {Romer}, {Rykoff}, {Sanchez},
  {Scarpine}, {Schubnell}, {Smith}, {Soares-Santos}, {Sobreira}, {Tarle},
  {Vikram}, {Walker}, \& {Zhang}}]{Etherington}
{Etherington}, J., {Thomas}, D., {Maraston}, C., {et~al.} 2017,
  \href{http://dx.doi.org/10.1093/mnras/stw3069}{\JournalTitle{\mnras}, 466,
  228}

\bibitem[{{Feeney} {et~al.}(2018){Feeney}, {Peiris}, {Williamson}, {Nissanke},
  {Mortlock}, {Alsing}, \& {Scolnic}}]{2018arXiv180203404F}
{Feeney}, S.~M., {Peiris}, H.~V., {Williamson}, A.~R., {et~al.} 2018,
  \JournalTitle{ArXiv e-prints},
  \href{http://arxiv.org/abs/1802.03404}{{\sffamily arXiv:1802.03404}}

\bibitem[{{Fishbach} {et~al.}(2018){Fishbach}, {Gray}, {Maga{\~n}a Hernandez},
  {Qi}, {Sur}, {members of the LIGO Scientific Collaboration}, \& {the Virgo
  Collaboration}}]{fishbach}
{Fishbach}, M., {Gray}, R., {Maga{\~n}a Hernandez}, I., {et~al.} 2018,
  \JournalTitle{ArXiv e-prints},
  \href{http://arxiv.org/abs/1807.05667}{{\sffamily arXiv:1807.05667}}

\bibitem[{{Flaugher} {et~al.}(2015){Flaugher}, {Diehl}, {Honscheid}, {Abbott},
  {Alvarez}, {Angstadt}, {Annis}, {Antonik}, {Ballester}, {Beaufore},
  {Bernstein}, {Bernstein}, {Bigelow}, {Bonati}, {Boprie}, {Brooks},
  {Buckley-Geer}, {Campa}, {Cardiel-Sas}, {Castander}, {Castilla}, {Cease},
  {Cela-Ruiz}, {Chappa}, {Chi}, {Cooper}, {da Costa}, {Dede}, {Derylo},
  {DePoy}, {de Vicente}, {Doel}, {Drlica-Wagner}, {Eiting}, {Elliott}, {Emes},
  {Estrada}, {Fausti Neto}, {Finley}, {Flores}, {Frieman}, {Gerdes},
  {Gladders}, {Gregory}, {Gutierrez}, {Hao}, {Holland}, {Holm}, {Huffman},
  {Jackson}, {James}, {Jonas}, {Karcher}, {Karliner}, {Kent}, {Kessler},
  {Kozlovsky}, {Kron}, {Kubik}, {Kuehn}, {Kuhlmann}, {Kuk}, {Lahav}, {Lathrop},
  {Lee}, {Levi}, {Lewis}, {Li}, {Mandrichenko}, {Marshall}, {Martinez},
  {Merritt}, {Miquel}, {Mu{\~n}oz}, {Neilsen}, {Nichol}, {Nord}, {Ogando},
  {Olsen}, {Palaio}, {Patton}, {Peoples}, {Plazas}, {Rauch}, {Reil}, {Rheault},
  {Roe}, {Rogers}, {Roodman}, {Sanchez}, {Scarpine}, {Schindler}, {Schmidt},
  {Schmitt}, {Schubnell}, {Schultz}, {Schurter}, {Scott}, {Serrano}, {Shaw},
  {Smith}, {Soares-Santos}, {Stefanik}, {Stuermer}, {Suchyta}, {Sypniewski},
  {Tarle}, {Thaler}, {Tighe}, {Tran}, {Tucker}, {Walker}, {Wang}, {Watson},
  {Weaverdyck}, {Wester}, {Woods}, {Yanny}, \& {The DES
  Collaboration}}]{flaugher}
{Flaugher}, B., {Diehl}, H.~T., {Honscheid}, K., {et~al.} 2015,
  \href{http://dx.doi.org/10.1088/0004-6256/150/5/150}{\JournalTitle{\aj}, 150,
  150}

\bibitem[{{Freedman}(2017)}]{2017NatAs...1E.169F}
{Freedman}, W.~L. 2017,
  \href{http://dx.doi.org/10.1038/s41550-017-0169}{\JournalTitle{Nature
  Astronomy}, 1, 0169}

\bibitem[{{Garc{\'{\i}}a-Bellido}(2017)}]{pbh}
{Garc{\'{\i}}a-Bellido}, J. 2017,
  \href{http://dx.doi.org/10.1088/1742-6596/840/1/012032}{in Journal of Physics
  Conference Series, Vol. 840, Journal of Physics Conference Series}, 012032

\bibitem[{{G{\'o}rski} {et~al.}(2005){G{\'o}rski}, {Hivon}, {Banday},
  {Wandelt}, {Hansen}, {Reinecke}, \& {Bartelmann}}]{healpix}
{G{\'o}rski}, K.~M., {Hivon}, E., {Banday}, A.~J., {et~al.} 2005,
  \href{http://dx.doi.org/10.1086/427976}{\JournalTitle{\apj}, 622, 759}

\bibitem[{{Gschwend} {et~al.}(2018){Gschwend}, {Rossel}, {Ogando}, {Neto},
  {Maia}, {da Costa}, {Lima}, {Pellegrini}, {Campisano}, {Singulani}, {Adean},
  {Benoist}, {Aguena}, {Carrasco Kind}, {Davis}, {de Vicente}, {Hartley},
  {Hoyle}, {Palmese}, {Sadeh}, {Abbott}, {Abdalla}, {Allam}, {Annis}, {Asorey},
  {Brooks}, {Calcino}, {Carollo}, {Castander}, {D'Andrea}, {Desai}, {Evrard},
  {Fosalba}, {Frieman}, {Garc{\'{\i}}a-Bellido}, {Glazebrook}, {Gerdes},
  {Gruendl}, {Gutierrez}, {Hinton}, {Hollowood}, {Honscheid}, {Hoormann},
  {James}, {Kuehn}, {Kuropatkin}, {Lahav}, {Lewis}, {Lidman}, {Lin},
  {Macaulay}, {Marshall}, {Melchior}, {Miquel}, {M{\"o}ller}, {Plazas},
  {Sanchez}, {Santiago}, {Scarpine}, {Schindler}, {Sevilla-Noarbe}, {Smith},
  {Sobreira}, {Sommer}, {Suchyta}, {Swanson}, {Tarle}, {Tucker}, {Tucker},
  {Uddin}, \& {Walker}}]{julia}
{Gschwend}, J., {Rossel}, A.~C., {Ogando}, R.~L.~C., {et~al.} 2018,
  \href{http://dx.doi.org/10.1016/j.ascom.2018.08.008}{\JournalTitle{Astronomy
  and Computing}, 25, 58}

\bibitem[{{Hartley} {et~al.}(2013){Hartley}, {Almaini}, {Mortlock},
  {Conselice}, {Gr{\"u}tzbauch}, {Simpson}, {Bradshaw}, {Chuter}, {Foucaud},
  {Cirasuolo}, {Dunlop}, {McLure}, \& {Pearce}}]{Hartley13}
{Hartley}, W.~G., {Almaini}, O., {Mortlock}, A., {et~al.} 2013,
  \href{http://dx.doi.org/10.1093/mnras/stt383}{\JournalTitle{\mnras}, 431,
  3045}

\bibitem[{{Holz} \& {Hughes}(2005)}]{2005ApJ...629...15H}
{Holz}, D.~E., \& {Hughes}, S.~A. 2005,
  \href{http://dx.doi.org/10.1086/431341}{\JournalTitle{\apj}, 629, 15}

\bibitem[{{Huchra} {et~al.}(2012){Huchra}, {Macri}, {Masters}, {Jarrett},
  {Berlind}, {Calkins}, {Crook}, {Cutri}, {Erdo{\v g}du}, {Falco}, {George},
  {Hutcheson}, {Lahav}, {Mader}, {Mink}, {Martimbeau}, {Schneider},
  {Skrutskie}, {Tokarz}, \& {Westover}}]{2012ApJS..199...26H}
{Huchra}, J.~P., {Macri}, L.~M., {Masters}, K.~L., {et~al.} 2012,
  \href{http://dx.doi.org/10.1088/0067-0049/199/2/26}{\JournalTitle{\apjs},
  199, 26}

\bibitem[{{Ilbert} {et~al.}(2006){Ilbert}, {Arnouts}, {McCracken},
  {Bolzonella}, {Bertin}, {Le F{\`e}vre}, {Mellier}, {Zamorani}, {Pell{\`o}},
  {Iovino}, {Tresse}, {Le Brun}, {Bottini}, {Garilli}, {Maccagni}, \&
  {Picat}}]{ilbertlephare}
{Ilbert}, O., {Arnouts}, S., {McCracken}, H.~J., {et~al.} 2006,
  \href{http://dx.doi.org/10.1051/0004-6361:20065138}{\JournalTitle{\aap}, 457,
  841}

\bibitem[{{Jones} {et~al.}(2009){Jones}, {Read}, {Saunders}, {Colless},
  {Jarrett}, {Parker}, {Fairall}, {Mauch}, {Sadler}, {Watson}, {Burton},
  {Campbell}, {Cass}, {Croom}, {Dawe}, {Fiegert}, {Frankcombe}, {Hartley},
  {Huchra}, {James}, {Kirby}, {Lahav}, {Lucey}, {Mamon}, {Moore}, {Peterson},
  {Prior}, {Proust}, {Russell}, {Safouris}, {Wakamatsu}, {Westra}, \&
  {Williams}}]{2009MNRAS.399..683J}
{Jones}, D.~H., {Read}, M.~A., {Saunders}, W., {et~al.} 2009,
  \href{http://dx.doi.org/10.1111/j.1365-2966.2009.15338.x}{\JournalTitle{\mnras},
  399, 683}

\bibitem[{{Leistedt} {et~al.}(2016){Leistedt}, {Peiris}, {Elsner},
  {Benoit-L{\'e}vy}, {Amara}, {Bauer}, {Becker}, {Bonnett}, {Bruderer},
  {Busha}, {Carrasco Kind}, {Chang}, {Crocce}, {da Costa}, {Gaztanaga}, {Huff},
  {Lahav}, {Palmese}, {Percival}, {Refregier}, {Ross}, {Rozo}, {Rykoff},
  {S{\'a}nchez}, {Sadeh}, {Sevilla-Noarbe}, {Sobreira}, {Suchyta}, {Swanson},
  {Wechsler}, {Abdalla}, {Allam}, {Banerji}, {Bernstein}, {Bernstein},
  {Bertin}, {Bridle}, {Brooks}, {Buckley-Geer}, {Burke}, {Capozzi}, {Carnero
  Rosell}, {Carretero}, {Cunha}, {D'Andrea}, {DePoy}, {Desai}, {Diehl}, {Doel},
  {Eifler}, {Evrard}, {Fausti Neto}, {Flaugher}, {Fosalba}, {Frieman},
  {Gerdes}, {Gruen}, {Gruendl}, {Gutierrez}, {Honscheid}, {James}, {Jarvis},
  {Kent}, {Kuehn}, {Kuropatkin}, {Li}, {Lima}, {Maia}, {March}, {Marshall},
  {Martini}, {Melchior}, {Miller}, {Miquel}, {Nichol}, {Nord}, {Ogando},
  {Plazas}, {Reil}, {Romer}, {Roodman}, {Sanchez}, {Santiago}, {Scarpine},
  {Schubnell}, {Smith}, {Soares-Santos}, {Tarle}, {Thaler}, {Thomas}, {Vikram},
  {Walker}, {Wester}, {Zhang}, \& {Zuntz}}]{boris}
{Leistedt}, B., {Peiris}, H.~V., {Elsner}, F., {et~al.} 2016,
  \href{http://dx.doi.org/10.3847/0067-0049/226/2/24}{\JournalTitle{\apjs},
  226, 24}

\bibitem[{{LIGO Scientific Collaboration} \& {Virgo
  Collaboration}(2017)}]{gcn_skymap}
{LIGO Scientific Collaboration}, \& {Virgo Collaboration}. 2017,
  \JournalTitle{GCN 21934}

\bibitem[{{LIGO Scientific Collaboration} {et~al.}(2017){LIGO Scientific
  Collaboration}, {Virgo Collaboration}, {GBM}, {INTEGRAL}, {IceCube
  Collaboration}, {AstroSat Cadmium Zinc Telluride Imager Team}, {IPN
  Collaboration}, {The Insight-Hxmt Collaboration}, {ANTARES Collaboration},
  {The Swift Collaboration}, {AGILE Team}, {The 1M2H Team}, {The Dark Energy
  Camera GW-EM Collaboration}, {the DES Collaboration}, {The DLT40
  Collaboration}, {GRAWITA}, {:}, {GRAvitational Wave Inaf TeAm}, {The Fermi
  Large Area Telescope Collaboration}, {ATCA}, {:}, {Telescope Compact Array},
  {ASKAP}, {:}, {SKA Pathfinder}, {Las Cumbres Observatory Group}, {OzGrav},
  {DWF}, {AST3}, {CAASTRO Collaborations}, {The VINROUGE Collaboration},
  {MASTER Collaboration}, {J-GEM}, {GROWTH}, {JAGWAR}, {NRAO}, {TTU-NRAO},
  {NuSTAR Collaborations}, {Pan-STARRS}, {The MAXI Team}, {Consortium}, {KU
  Collaboration}, {Optical Telescope}, {ePESSTO}, {GROND}, {Tech University},
  {SALT Group}, {TOROS}, {:}, {Transient Robotic Observatory of the South
  Collaboration}, {The BOOTES Collaboration}, {MWA}, {:}, {Widefield Array},
  {The CALET Collaboration}, {IKI-GW Follow-up Collaboration},
  {H.~E.~S.~S.~Collaboration}, {LOFAR Collaboration}, {LWA}, {:}, {Wavelength
  Array}, {HAWC Collaboration}, {The Pierre Auger Collaboration}, {ALMA
  Collaboration}, {Euro VLBI Team}, {Pi of the Sky Collaboration}, {The Chandra
  Team at McGill University}, {DFN}, {:}, {Fireball Network}, {ATLAS}, {Time
  Resolution Universe Survey}, {RIMAS}, {RATIR}, \& {South
  Africa/MeerKAT}}]{MMApaper}
{LIGO Scientific Collaboration}, {Virgo Collaboration}, {GBM}, F., {et~al.}
  2017, \JournalTitle{ArXiv e-prints},
  \href{http://arxiv.org/abs/1710.05833}{{\sffamily arXiv:1710.05833
  [astro-ph.HE]}}

\bibitem[{{Lima} {et~al.}(2008){Lima}, {Cunha}, {Oyaizu}, {Frieman}, {Lin}, \&
  {Sheldon}}]{lima}
{Lima}, M., {Cunha}, C.~E., {Oyaizu}, H., {et~al.} 2008,
  \href{http://dx.doi.org/10.1111/j.1365-2966.2008.13510.x}{\JournalTitle{\mnras},
  390, 118}

\bibitem[{{Lipunov} {et~al.}(2017){Lipunov}, {Gorbovskoy}, {Kornilov},
  {.~Tyurina}, {Balanutsa}, {Kuznetsov}, {Vlasenko}, {Kuvshinov}, {Gorbunov},
  {Buckley}, {Krylov}, {Podesta}, {Lopez}, {Podesta}, {Levato}, {Saffe},
  {Mallamachi}, {Potter}, {Budnev}, {Gress}, {Ishmuhametova}, {Vladimirov},
  {Zimnukhov}, {Yurkov}, {Sergienko}, {Gabovich}, {Rebolo}, {Serra-Ricart},
  {Israelyan}, {Chazov}, {Wang}, {Tlatov}, \& {Panchenko}}]{lipunov}
{Lipunov}, V.~M., {Gorbovskoy}, E., {Kornilov}, V.~G., {et~al.} 2017,
  \href{http://dx.doi.org/10.3847/2041-8213/aa92c0}{\JournalTitle{\apjl}, 850,
  L1}

\bibitem[{{LSST Dark Energy Science Collaboration}(2012)}]{2012arXiv1211.0310L}
{LSST Dark Energy Science Collaboration}. 2012, \JournalTitle{ArXiv e-prints},
  \href{http://arxiv.org/abs/1211.0310}{{\sffamily arXiv:1211.0310
  [astro-ph.CO]}}

\bibitem[{{Macaulay} {et~al.}(2018){Macaulay}, {Nichol}, {Bacon}, {Brout},
  {Davis}, {Zhang}, {Bassett}, {Scolnic}, {M{\"o}ller}, {D'Andrea}, {Hinton},
  {Kessler}, {Kim}, {Lasker}, {Lidman}, {Sako}, {Smith}, {Sullivan}, {Abbott},
  {Allam}, {Annis}, {Asorey}, {Avila}, {Bechtol}, {Brooks}, {Brown}, {Burke},
  {Calcino}, {Carnero Rosell}, {Carollo}, {Carrasco Kind}, {Carretero},
  {Castander}, {Collett}, {Crocce}, {Cunha}, {da Costa}, {Davis}, {De Vicente},
  {Diehl}, {Doel}, {Drlica-Wagner}, {Eifler}, {Estrada}, {Evrard}, {Finley},
  {Flaugher}, {Foley}, {Fosalba}, {Frieman}, {Galbany},
  {Garc{\'{\i}}a-Bellido}, {Gaztanaga}, {Glazebrook},
  {Gonz{\'a}lez-Gait{\'a}n}, {Gruen}, {Gruendl}, {Gschwend}, {Gutierrez},
  {Hartley}, {Hollowood}, {Honscheid}, {Hoormann}, {Hoyle}, {Huterer}, {Jain},
  {James}, {Jeltema}, {Kasai}, {Krause}, {Kuehn}, {Kuropatkin}, {Lahav},
  {Lewis}, {Li}, {Lima}, {Lin}, {Maia}, {Marshall}, {Martini}, {Miquel},
  {Nugent}, {Palmese}, {Pan}, {Plazas}, {Romer}, {Roodman}, {Sanchez},
  {Scarpine}, {Schindler}, {Schubnell}, {Serrano}, {Sevilla-Noarbe}, {Sharp},
  {Soares-Santos}, {Sobreira}, {Sommer}, {Suchyta}, {Swann}, {Swanson},
  {Tarle}, {Thomas}, {Thomas}, {Tucker}, {Uddin}, {Vikram}, {Walker}, \&
  {Wiseman}}]{2018arXiv181102376M}
{Macaulay}, E., {Nichol}, R.~C., {Bacon}, D., {et~al.} 2018,
  \JournalTitle{ArXiv e-prints},
  \href{http://arxiv.org/abs/1811.02376}{{\sffamily arXiv:1811.02376}}

\bibitem[{{MacLeod} \& {Hogan}(2008)}]{macleod}
{MacLeod}, C.~L., \& {Hogan}, C.~J. 2008,
  \href{http://dx.doi.org/10.1103/PhysRevD.77.043512}{\JournalTitle{\prd}, 77,
  043512}

\bibitem[{{Mandel} {et~al.}(2018){Mandel}, {Farr}, \& {Gair}}]{mandel}
{Mandel}, I., {Farr}, W.~M., \& {Gair}, J.~R. 2018, \JournalTitle{ArXiv
  e-prints}, \href{http://arxiv.org/abs/1809.02063}{{\sffamily arXiv:1809.02063
  [physics.data-an]}}

\bibitem[{{Mapelli} {et~al.}(2018){Mapelli}, {Giacobbo}, {Toffano},
  {Ripamonti}, {Bressan}, {Spera}, \& {Branchesi}}]{2018MNRAS.481.5324M}
{Mapelli}, M., {Giacobbo}, N., {Toffano}, M., {et~al.} 2018,
  \href{http://dx.doi.org/10.1093/mnras/sty2663}{\JournalTitle{\mnras}, 481,
  5324}

\bibitem[{{Morganson} {et~al.}(2018){Morganson}, {Gruendl}, {Menanteau},
  {Carrasco Kind}, {Chen}, {Daues}, {Drlica-Wagner}, {Friedel}, {Gower},
  {Johnson}, {Johnson}, {Kessler}, {Paz-Chinch{\'o}n}, {Petravick}, {Pond},
  {Yanny}, {Allam}, {Armstrong}, {Barkhouse}, {Bechtol}, {Benoit-L{\'e}vy},
  {Bernstein}, {Bertin}, {Buckley-Geer}, {Covarrubias}, {Desai}, {Diehl},
  {Goldstein}, {Gruen}, {Li}, {Lin}, {Marriner}, {Mohr}, {Neilsen}, {Ngeow},
  {Paech}, {Rykoff}, {Sako}, {Sevilla-Noarbe}, {Sheldon}, {Sobreira}, {Tucker},
  {Wester}, \& {DES Collaboration}}]{2018PASP..130g4501M}
{Morganson}, E., {Gruendl}, R.~A., {Menanteau}, F., {et~al.} 2018,
  \href{http://dx.doi.org/10.1088/1538-3873/aab4ef}{\JournalTitle{\pasp}, 130,
  074501}

\bibitem[{{Mortlock} {et~al.}(2018){Mortlock}, {Feeney}, {Peiris},
  {Williamson}, \& {Nissanke}}]{2018arXiv181111723M}
{Mortlock}, D.~J., {Feeney}, S.~M., {Peiris}, H.~V., {Williamson}, A.~R., \&
  {Nissanke}, S.~M. 2018, \JournalTitle{arXiv e-prints}, arXiv:1811.11723

\bibitem[{{M{\"o}rtsell} \& {Dhawan}(2018)}]{2018JCAP...09..025M}
{M{\"o}rtsell}, E., \& {Dhawan}, S. 2018,
  \href{http://dx.doi.org/10.1088/1475-7516/2018/09/025}{\JournalTitle{\jcap},
  9, 025}

\bibitem[{{Nair} {et~al.}(2018){Nair}, {Bose}, \& {Deep Saini}}]{nair}
{Nair}, R., {Bose}, S., \& {Deep Saini}, T. 2018, \JournalTitle{ArXiv
  e-prints}, \href{http://arxiv.org/abs/1804.06085}{{\sffamily
  arXiv:1804.06085}}

\bibitem[{{Nishizawa}(2017)}]{2017PhRvD..96j1303N}
{Nishizawa}, A. 2017,
  \href{http://dx.doi.org/10.1103/PhysRevD.96.101303}{\JournalTitle{\prd}, 96,
  101303}

\bibitem[{{Nissanke} {et~al.}(2013){Nissanke}, {Holz}, {Dalal}, {Hughes},
  {Sievers}, \& {Hirata}}]{2013arXiv1307.2638N}
{Nissanke}, S., {Holz}, D.~E., {Dalal}, N., {et~al.} 2013, \JournalTitle{ArXiv
  e-prints}, \href{http://arxiv.org/abs/1307.2638}{{\sffamily arXiv:1307.2638
  [astro-ph.CO]}}

\bibitem[{{Nissanke} {et~al.}(2010){Nissanke}, {Holz}, {Hughes}, {Dalal}, \&
  {Sievers}}]{2010ApJ...725..496N}
{Nissanke}, S., {Holz}, D.~E., {Hughes}, S.~A., {Dalal}, N., \& {Sievers},
  J.~L. 2010,
  \href{http://dx.doi.org/10.1088/0004-637X/725/1/496}{\JournalTitle{\apj},
  725, 496}

\bibitem[{{Palmese} {et~al.}(2019){Palmese}, {Annis}, {Burgad}, {Lin},
  {Pereira}, \& M.~Soares-Santos$^{3,1}$}]{palmese18}
{Palmese}, A., {Annis}, T.~J., {Burgad}, J.~C., {et~al.} 2019, \JournalTitle{in
  prep.}

\bibitem[{{Palmese} {et~al.}(2016){Palmese}, {Lahav}, {Banerji}, {Gruen},
  {Jouvel}, {Melchior}, {Aleksi{\'c}}, {Annis}, {Diehl}, {Hartley}, {Jeltema},
  {Romer}, {Rozo}, {Rykoff}, {Seitz}, {Suchyta}, {Zhang}, {Abbott}, {Abdalla},
  {Allam}, {Benoit-L{\'e}vy}, {Bertin}, {Brooks}, {Buckley-Geer}, {Burke},
  {Capozzi}, {Carnero Rosell}, {Carrasco Kind}, {Carretero}, {Crocce}, {Cunha},
  {D'Andrea}, {da Costa}, {Desai}, {Dietrich}, {Doel}, {Estrada}, {Evrard},
  {Flaugher}, {Frieman}, {Gerdes}, {Goldstein}, {Gruendl}, {Gutierrez},
  {Honscheid}, {James}, {Kuehn}, {Kuropatkin}, {Li}, {Lima}, {Maia},
  {Marshall}, {Miller}, {Miquel}, {Nord}, {Ogando}, {Plazas}, {Roodman},
  {Sanchez}, {Scarpine}, {Sevilla-Noarbe}, {Smith}, {Soares-Santos},
  {Sobreira}, {Swanson}, {Tarle}, {Thomas}, {Tucker}, \& {Vikram}}]{palmese16}
{Palmese}, A., {Lahav}, O., {Banerji}, M., {et~al.} 2016,
  \href{http://dx.doi.org/10.1093/mnras/stw2062}{\JournalTitle{\mnras}, 463,
  1486}

\bibitem[{{Palmese} {et~al.}(2017){Palmese}, {Hartley}, {Tarsitano},
  {Conselice}, {Lahav}, {Allam}, {Annis}, {Lin}, {Soares-Santos}, {Tucker},
  {Brout}, {Banerji}, {Bechtol}, {Diehl}, {Fruchter}, {Garc{\'{\i}}a-Bellido},
  {Herner}, {Levan}, {Li}, {Lidman}, {Misra}, {Sako}, {Scolnic}, {Smith},
  {Abbott}, {Abdalla}, {Benoit-L{\'e}vy}, {Bertin}, {Brooks}, {Buckley-Geer},
  {Carnero Rosell}, {Carrasco Kind}, {Carretero}, {Castander}, {Cunha},
  {D'Andrea}, {da Costa}, {Davis}, {DePoy}, {Desai}, {Dietrich}, {Doel},
  {Drlica-Wagner}, {Eifler}, {Evrard}, {Flaugher}, {Fosalba}, {Frieman},
  {Gaztanaga}, {Gerdes}, {Giannantonio}, {Gruen}, {Gruendl}, {Gschwend},
  {Gutierrez}, {Honscheid}, {Jain}, {James}, {Jeltema}, {Johnson}, {Johnson},
  {Krause}, {Kron}, {Kuehn}, {Kuhlmann}, {Kuropatkin}, {Lima}, {Maia}, {March},
  {Marshall}, {McMahon}, {Menanteau}, {Miller}, {Miquel}, {Neilsen}, {Ogando},
  {Plazas}, {Reil}, {Romer}, {Sanchez}, {Schindler}, {Smith}, {Sobreira},
  {Suchyta}, {Swanson}, {Tarle}, {Thomas}, {Thomas}, {Walker}, {Weller},
  {Zhang}, \& {Zuntz}}]{palmese}
{Palmese}, A., {Hartley}, W., {Tarsitano}, F., {et~al.} 2017,
  \href{http://dx.doi.org/10.3847/2041-8213/aa9660}{\JournalTitle{\apjl}, 849,
  L34}

\bibitem[{{Planck Collaboration} {et~al.}(2018){Planck Collaboration},
  {Aghanim}, {Akrami}, {Ashdown}, {Aumont}, {Baccigalupi}, {Ballardini},
  {Banday}, {Barreiro}, {Bartolo}, {Basak}, {Battye}, {Benabed}, {Bernard},
  {Bersanelli}, {Bielewicz}, {Bock}, {Bond}, {Borrill}, {Bouchet}, {Boulanger},
  {Bucher}, {Burigana}, {Butler}, {Calabrese}, {Cardoso}, {Carron},
  {Challinor}, {Chiang}, {Chluba}, {Colombo}, {Combet}, {Contreras}, {Crill},
  {Cuttaia}, {de Bernardis}, {de Zotti}, {Delabrouille}, {Delouis}, {Di
  Valentino}, {Diego}, {Dor{\'e}}, {Douspis}, {Ducout}, {Dupac}, {Dusini},
  {Efstathiou}, {Elsner}, {En{\ss}lin}, {Eriksen}, {Fantaye}, {Farhang},
  {Fergusson}, {Fernandez-Cobos}, {Finelli}, {Forastieri}, {Frailis},
  {Franceschi}, {Frolov}, {Galeotta}, {Galli}, {Ganga}, {G{\'e}nova-Santos},
  {Gerbino}, {Ghosh}, {Gonz{\'a}lez-Nuevo}, {G{\'o}rski}, {Gratton},
  {Gruppuso}, {Gudmundsson}, {Hamann}, {Handley}, {Herranz}, {Hivon}, {Huang},
  {Jaffe}, {Jones}, {Karakci}, {Keih{\"a}nen}, {Keskitalo}, {Kiiveri}, {Kim},
  {Kisner}, {Knox}, {Krachmalnicoff}, {Kunz}, {Kurki-Suonio}, {Lagache},
  {Lamarre}, {Lasenby}, {Lattanzi}, {Lawrence}, {Le Jeune}, {Lemos},
  {Lesgourgues}, {Levrier}, {Lewis}, {Liguori}, {Lilje}, {Lilley}, {Lindholm},
  {L{\'o}pez-Caniego}, {Lubin}, {Ma}, {Mac{\'{\i}}as-P{\'e}rez}, {Maggio},
  {Maino}, {Mandolesi}, {Mangilli}, {Marcos-Caballero}, {Maris}, {Martin},
  {Martinelli}, {Mart{\'{\i}}nez-Gonz{\'a}lez}, {Matarrese}, {Mauri}, {McEwen},
  {Meinhold}, {Melchiorri}, {Mennella}, {Migliaccio}, {Millea}, {Mitra},
  {Miville-Desch{\^e}nes}, {Molinari}, {Montier}, {Morgante}, {Moss}, {Natoli},
  {N{\o}rgaard-Nielsen}, {Pagano}, {Paoletti}, {Partridge}, {Patanchon},
  {Peiris}, {Perrotta}, {Pettorino}, {Piacentini}, {Polastri}, {Polenta},
  {Puget}, {Rachen}, {Reinecke}, {Remazeilles}, {Renzi}, {Rocha}, {Rosset},
  {Roudier}, {Rubi{\~n}o-Mart{\'{\i}}n}, {Ruiz-Granados}, {Salvati}, {Sandri},
  {Savelainen}, {Scott}, {Shellard}, {Sirignano}, {Sirri}, {Spencer},
  {Sunyaev}, {Suur-Uski}, {Tauber}, {Tavagnacco}, {Tenti}, {Toffolatti},
  {Tomasi}, {Trombetti}, {Valenziano}, {Valiviita}, {Van Tent}, {Vibert},
  {Vielva}, {Villa}, {Vittorio}, {Wandelt}, {Wehus}, {White}, {White},
  {Zacchei}, \& {Zonca}}]{planck18}
{Planck Collaboration}, {Aghanim}, N., {Akrami}, Y., {et~al.} 2018,
  \JournalTitle{ArXiv e-prints},
  \href{http://arxiv.org/abs/1807.06209}{{\sffamily arXiv:1807.06209}}

\bibitem[{{Pozzetti} {et~al.}(2010){Pozzetti}, {Bolzonella}, {Zucca},
  {Zamorani}, {Lilly}, {Renzini}, {Moresco}, {Mignoli}, {Cassata}, {Tasca},
  {Lamareille}, {Maier}, {Meneux}, {Halliday}, {Oesch}, {Vergani}, {Caputi},
  {Kova{\v c}}, {Cimatti}, {Cucciati}, {Iovino}, {Peng}, {Carollo}, {Contini},
  {Kneib}, {Le F{\'e}vre}, {Mainieri}, {Scodeggio}, {Bardelli}, {Bongiorno},
  {Coppa}, {de la Torre}, {de Ravel}, {Franzetti}, {Garilli}, {Kampczyk},
  {Knobel}, {Le Borgne}, {Le Brun}, {Pell{\`o}}, {Perez Montero},
  {Ricciardelli}, {Silverman}, {Tanaka}, {Tresse}, {Abbas}, {Bottini}, {Cappi},
  {Guzzo}, {Koekemoer}, {Leauthaud}, {Maccagni}, {Marinoni}, {McCracken},
  {Memeo}, {Porciani}, {Scaramella}, {Scarlata}, \& {Scoville}}]{Pozzetti10}
{Pozzetti}, L., {Bolzonella}, M., {Zucca}, E., {et~al.} 2010,
  \href{http://dx.doi.org/10.1051/0004-6361/200913020}{\JournalTitle{\aap},
  523, A13}

\bibitem[{{Riess} {et~al.}(2016){Riess}, {Macri}, {Hoffmann}, {Scolnic},
  {Casertano}, {Filippenko}, {Tucker}, {Reid}, {Jones}, {Silverman},
  {Chornock}, {Challis}, {Yuan}, {Brown}, \& {Foley}}]{shoes}
{Riess}, A.~G., {Macri}, L.~M., {Hoffmann}, S.~L., {et~al.} 2016,
  \href{http://dx.doi.org/10.3847/0004-637X/826/1/56}{\JournalTitle{\apj}, 826,
  56}

\bibitem[{{Riess} {et~al.}(2018){Riess}, {Casertano}, {Yuan}, {Macri},
  {Bucciarelli}, {Lattanzi}, {MacKenty}, {Bowers}, {Zheng}, {Filippenko},
  {Huang}, \& {Anderson}}]{riess18}
{Riess}, A.~G., {Casertano}, S., {Yuan}, W., {et~al.} 2018,
  \href{http://dx.doi.org/10.3847/1538-4357/aac82e}{\JournalTitle{\apj}, 861,
  126}

\bibitem[{{Sadeh} {et~al.}(2016){Sadeh}, {Abdalla}, \& {Lahav}}]{annz2}
{Sadeh}, I., {Abdalla}, F.~B., \& {Lahav}, O. 2016,
  \href{http://dx.doi.org/10.1088/1538-3873/128/968/104502}{\JournalTitle{\pasp},
  128, 104502}

\bibitem[{{Schutz}(1986)}]{schutz}
{Schutz}, B.~F. 1986,
  \href{http://dx.doi.org/10.1038/323310a0}{\JournalTitle{\nat}, 323, 310}

\bibitem[{{Shectman} {et~al.}(1996){Shectman}, {Landy}, {Oemler}, {Tucker},
  {Lin}, {Kirshner}, \& {Schechter}}]{1996ApJ...470..172S}
{Shectman}, S.~A., {Landy}, S.~D., {Oemler}, A., {et~al.} 1996,
  \href{http://dx.doi.org/10.1086/177858}{\JournalTitle{\apj}, 470, 172}

\bibitem[{{Singer} {et~al.}(2016){Singer}, {Chen}, {Holz}, {Farr}, {Price},
  {Raymond}, {Cenko}, {Gehrels}, {Cannizzo}, {Kasliwal}, {Nissanke},
  {Coughlin}, {Farr}, {Urban}, {Vitale}, {Veitch}, {Graff}, {Berry},
  {Mohapatra}, \& {Mandel}}]{Singer}
{Singer}, L.~P., {Chen}, H.-Y., {Holz}, D.~E., {et~al.} 2016,
  \href{http://dx.doi.org/10.3847/0067-0049/226/1/10}{\JournalTitle{\apjs},
  226, 10}

\bibitem[{{Soares-Santos} {et~al.}(2016){Soares-Santos}, {Kessler}, {Berger},
  {Annis}, {Brout}, {Buckley-Geer}, {Chen}, {Cowperthwaite}, {Diehl}, {Doctor},
  {Drlica-Wagner}, {Farr}, {Finley}, {Flaugher}, {Foley}, {Frieman}, {Gruendl},
  {Herner}, {Holz}, {Lin}, {Marriner}, {Neilsen}, {Rest}, {Sako}, {Scolnic},
  {Sobreira}, {Walker}, {Wester}, {Yanny}, {Abbott}, {Abdalla}, {Allam},
  {Armstrong}, {Banerji}, {Benoit-L{\'e}vy}, {Bernstein}, {Bertin}, {Brown},
  {Burke}, {Capozzi}, {Carnero Rosell}, {Carrasco Kind}, {Carretero},
  {Castander}, {Cenko}, {Chornock}, {Crocce}, {D'Andrea}, {da Costa}, {Desai},
  {Dietrich}, {Drout}, {Eifler}, {Estrada}, {Evrard}, {Fairhurst}, {Fernandez},
  {Fischer}, {Fong}, {Fosalba}, {Fox}, {Fryer}, {Garcia-Bellido}, {Gaztanaga},
  {Gerdes}, {Goldstein}, {Gruen}, {Gutierrez}, {Honscheid}, {James},
  {Karliner}, {Kasen}, {Kent}, {Kuropatkin}, {Kuehn}, {Lahav}, {Li}, {Lima},
  {Maia}, {Margutti}, {Martini}, {Matheson}, {McMahon}, {Metzger}, {Miller},
  {Miquel}, {Mohr}, {Nichol}, {Nord}, {Ogando}, {Peoples}, {Plazas},
  {Quataert}, {Romer}, {Roodman}, {Rykoff}, {Sanchez}, {Scarpine}, {Schindler},
  {Schubnell}, {Sevilla-Noarbe}, {Sheldon}, {Smith}, {Smith}, {Smith},
  {Stebbins}, {Sutton}, {Swanson}, {Tarle}, {Thaler}, {Thomas}, {Tucker},
  {Vikram}, {Wechsler}, {Weller}, \& {DES Collaboration}}]{2016ApJ...823L..33S}
{Soares-Santos}, M., {Kessler}, R., {Berger}, E., {et~al.} 2016,
  \href{http://dx.doi.org/10.3847/2041-8205/823/2/L33}{\JournalTitle{\apjl},
  823, L33}

\bibitem[{{Soares-Santos} {et~al.}(2017){Soares-Santos}, {Holz}, {Annis},
  {Chornock}, {Herner}, {Berger}, {Brout}, {Chen}, {Kessler}, {Sako}, {Allam},
  {Tucker}, {Butler}, {Palmese}, {Doctor}, {Diehl}, {Frieman}, {Yanny}, {Lin},
  {Scolnic}, {Cowperthwaite}, {Neilsen}, {Marriner}, {Kuropatkin}, {Hartley},
  {Paz-Chinch{\'o}n}, {Alexander}, {Balbinot}, {Blanchard}, {Brown}, {Carlin},
  {Conselice}, {Cook}, {Drlica-Wagner}, {Drout}, {Durret}, {Eftekhari}, {Farr},
  {Finley}, {Foley}, {Fong}, {Fryer}, {Garc{\'{\i}}a-Bellido}, {Gill},
  {Gruendl}, {Hanna}, {Kasen}, {Li}, {Lopes}, {Louren{\c c}o}, {Margutti},
  {Marshall}, {Matheson}, {Medina}, {Metzger}, {Mu{\~n}oz}, {Muir}, {Nicholl},
  {Quataert}, {Rest}, {Sauseda}, {Schlegel}, {Secco}, {Sobreira}, {Stebbins},
  {Villar}, {Walker}, {Wester}, {Williams}, {Zenteno}, {Zhang}, {Abbott},
  {Abdalla}, {Banerji}, {Bechtol}, {Benoit-L{\'e}vy}, {Bertin}, {Brooks},
  {Buckley-Geer}, {Burke}, {Carnero Rosell}, {Carrasco Kind}, {Carretero},
  {Castander}, {Crocce}, {Cunha}, {D'Andrea}, {da Costa}, {Davis}, {Desai},
  {Dietrich}, {Doel}, {Eifler}, {Fernandez}, {Flaugher}, {Fosalba},
  {Gaztanaga}, {Gerdes}, {Giannantonio}, {Goldstein}, {Gruen}, {Gschwend},
  {Gutierrez}, {Honscheid}, {Jain}, {James}, {Jeltema}, {Johnson}, {Johnson},
  {Kent}, {Krause}, {Kron}, {Kuehn}, {Kuhlmann}, {Lahav}, {Lima}, {Maia},
  {March}, {McMahon}, {Menanteau}, {Miquel}, {Mohr}, {Nichol}, {Nord},
  {Ogando}, {Petravick}, {Plazas}, {Romer}, {Roodman}, {Rykoff}, {Sanchez},
  {Scarpine}, {Schubnell}, {Sevilla-Noarbe}, {Smith}, {Smith}, {Suchyta},
  {Swanson}, {Tarle}, {Thomas}, {Thomas}, {Troxel}, {Vikram}, {Wechsler}, \&
  {Weller}}]{marcelle17}
{Soares-Santos}, M., {Holz}, D.~E., {Annis}, J., {et~al.} 2017,
  \JournalTitle{ArXiv e-prints},
  \href{http://arxiv.org/abs/1710.05459}{{\sffamily arXiv:1710.05459
  [astro-ph.HE]}}

\bibitem[{{Tanvir} {et~al.}(2017){Tanvir}, {Levan},
  {Gonz{\'a}lez-Fern{\'a}ndez}, {Korobkin}, {Mandel}, {Rosswog}, {Hjorth},
  {D'Avanzo}, {Fruchter}, {Fryer}, {Kangas}, {Milvang-Jensen}, {Rosetti},
  {Steeghs}, {Wollaeger}, {Cano}, {Copperwheat}, {Covino}, {D'Elia}, {de Ugarte
  Postigo}, {Evans}, {Even}, {Fairhurst}, {Figuera Jaimes}, {Fontes}, {Fujii},
  {Fynbo}, {Gompertz}, {Greiner}, {Hodosan}, {Irwin}, {Jakobsson},
  {J{\o}rgensen}, {Kann}, {Lyman}, {Malesani}, {McMahon}, {Melandri},
  {O'Brien}, {Osborne}, {Palazzi}, {Perley}, {Pian}, {Piranomonte}, {Rabus},
  {Rol}, {Rowlinson}, {Schulze}, {Sutton}, {Th{\"o}ne}, {Ulaczyk}, {Watson},
  {Wiersema}, \& {Wijers}}]{tanvir}
{Tanvir}, N.~R., {Levan}, A.~J., {Gonz{\'a}lez-Fern{\'a}ndez}, C., {et~al.}
  2017, \href{http://dx.doi.org/10.3847/2041-8213/aa90b6}{\JournalTitle{\apjl},
  848, L27}

\bibitem[{{The Dark Energy Survey Collaboration}(2005)}]{descollaboration}
{The Dark Energy Survey Collaboration}. 2005, \JournalTitle{preprint
  (arXiv:astro-ph/0510346)},
  \href{http://arxiv.org/abs/astro-ph/0510346}{{\sffamily astro-ph/0510346}}

\bibitem[{{Valenti} {et~al.}(2017){Valenti}, {David}, {Sand}, {Yang},
  {Cappellaro}, {Tartaglia}, {Corsi}, {Jha}, {Reichart}, {Haislip}, \&
  {Kouprianov}}]{valenti}
{Valenti}, S., {David}, {Sand}, J., {et~al.} 2017,
  \href{http://dx.doi.org/10.3847/2041-8213/aa8edf}{\JournalTitle{\apjl}, 848,
  L24}

\bibitem[{{Vitale} \& {Chen}(2018)}]{2018PhRvL.121b1303V}
{Vitale}, S., \& {Chen}, H.-Y. 2018,
  \href{http://dx.doi.org/10.1103/PhysRevLett.121.021303}{\JournalTitle{Physical
  Review Letters}, 121, 021303}

\bibitem[{{Wechsler} \& {Tinker}(2018)}]{2018ARA&A..56..435W}
{Wechsler}, R.~H., \& {Tinker}, J.~L. 2018,
  \href{http://dx.doi.org/10.1146/annurev-astro-081817-051756}{\JournalTitle{\araa},
  56, 435}

\bibitem[{{Weigel} {et~al.}(2016){Weigel}, {Schawinski}, \&
  {Bruderer}}]{weigel}
{Weigel}, A.~K., {Schawinski}, K., \& {Bruderer}, C. 2016,
  \href{http://dx.doi.org/10.1093/mnras/stw756}{\JournalTitle{\mnras}, 459,
  2150}

\end{thebibliography}
\end{document}